\documentclass[a4paper,oneside,11pt]{article}
\usepackage{cprform}
\usepackage{amsmath,amstext,amsfonts,amsbsy,amssymb,amscd,bbm,epsfig,lscape}

\usepackage{epsfig,macros,cite}
\usepackage{amssymb}
\usepackage{subfigure}

\newcommand{\parbreak}{\vskip 0.15cm}

\newcommand{\req}[1]{Eq.~(\ref{#1})}

\newcommand{\rep}[1]{\cite{#1}}

\newcommand{\ci}[1]{\boldsymbol{#1}}

\newcommand{\Dslash}{\relax{\kern+.25em / \kern-.70em D}}

\newcommand{\Real}{\relax{\mathsf{\Gamma\kern-.35em R}}}
\newcommand{\Int}{\relax{\mathsf{Z\kern-.40em Z}}}

\newcommand{\mren}[1]{m_{{\rm R} #1}}
\newcommand{\muren}[1]{\mu_{{\rm R} #1}}
\newcommand{\obar}[1]{\kern3pt\overline{\kern-2pt #1\kern-0pt}\kern1pt}

\newcommand{\corrbar}[1]{\kern3pt\overline{\kern-2pt #1\kern-0pt}\kern1pt}

\newcommand{\oVApAVren}[1]{\kern3pt\overline{\kern-2pt #1\kern-0pt}\kern1pt_{\rm\scriptscriptstyle VA+AV;s}}

\newcommand{\Zm}{Z_{\rm m}}
\newcommand{\ZP}{Z_{\rm\scriptscriptstyle P}}

\newcommand{\ZA}{Z_{\rm\scriptscriptstyle A}}
\newcommand{\ZV}{Z_{\rm\scriptscriptstyle V}}

\newcommand{\zbar}{\kern3pt\overline{\kern-2pt Z\kern-0pt}\kern1pt}

\newcommand{\zbarVApAV}[1]{\kern3pt\overline{\kern-2pt Z\kern-0pt}\kern1pt_{\rm\scriptscriptstyle VA+AV #1}}

\newcommand{\icA}{c_{\rm\scriptscriptstyle A}}

\newcommand{\icV}{c_{\rm\scriptscriptstyle V}}
\newcommand{\ibP}{b_{\rm\scriptscriptstyle P}}
\newcommand{\ibA}{b_{\rm\scriptscriptstyle A}}

\newcommand{\ibV}{b_{\rm\scriptscriptstyle V}}
\newcommand{\ibm}{b_{\rm m}}

\newcommand{\cR}{{\cal R}}

\newcommand{\scrA}{{\rm\scriptscriptstyle A}}
\newcommand{\scrP}{{\rm\scriptscriptstyle P}}
\newcommand{\scrV}{{\rm\scriptscriptstyle V}}

\begin{document}
\bibliographystyle{mybibstyle}

%%%%%%%%%%%%%%%%%%%%%%%%%%%%%%%%%%%%%%%%%%%%%%%%
% TITLE

\begin{titlepage}

%%% Preprint numbers %%%

\vspace*{-30truemm}
\begin{flushright}
ROM2F/2007-03\\
MS-TP-07-1\\
CERN-PH-TH/2007-018\\
MKPH-T-07-03\\ 
TRINLAT-07/03\\
\vspace{5truemm}
{\large February 2007}
\end{flushright}
\vspace{5truemm}

%%% Title and authors %%%

\centerline{\Bigrm Flavour symmetry restoration and}
\centerline{\Bigrm kaon weak matrix elements in quenched twisted mass QCD}
\vskip 9 true mm
\begin{center}
\epsfig{figure=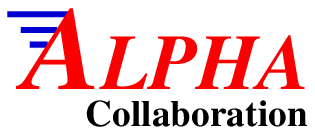, width=22 true mm}\\
\end{center}
\vskip -2 true mm
\centerline{\bigrm  P.~Dimopoulos$^a$, 
J.~Heitger$^b$,
F.~Palombi$^c$,
C.~Pena$^d$, 
S.~Sint$^e$ 
and A.~Vladikas$^a$}
\vskip 4 true mm
\centerline{\it $^a$ INFN, Sezione di Roma II}
\centerline{\it and Dipartimento di Fisica, Universit\`a di Roma ``Tor
  Vergata''}
\centerline{\it Via della Ricerca Scientifica 1, I-00133 Rome, Italy}
\vskip 3 true mm
\centerline{\it $^b$ Westf\"alische Wilhelms-Universit\"at M\"unster,
Institut f\"ur Theoretische Physik}
\centerline{\it Wilhelm-Klemm-Strasse 9, D-48149 M\"unster, Germany}
\vskip 3 true mm
\centerline{\it $^c$ Johannes Gutenberg Universit\"at, Institut f\"ur Kernphysik}
\centerline{\it Johann Joachim Becher-Weg 45, D-55099 Mainz, Germany}
\vskip 3 true mm
\centerline{\it $^d$ CERN, Physics Department, TH Division, CH-1211
  Geneva, Switzerland}
\vskip 3 true mm
\centerline{\it $^e$ School of Mathematics, Trinity College, Dublin 2, Ireland}
\vskip 10 true mm

%%% Abstract %%%

\thicktablerule
\vskip 3 true mm
\noindent{\tenbf Abstract}
\vskip 1 true mm
\noindent
{\tenrm We simulate two variants of quenched twisted mass QCD (tmQCD),
with degenerate Wilson quarks of masses equal to or heavier than
half the strange quark mass. We use Ward identities in order
to measure the twist angles of the theory and thus check the quality
of the tuning of mass parameters to a physics condition which stays
constant as the lattice spacing is varied. Flavour symmetry breaking in tmQCD
is studied in a framework of two fully twisted and two standard Wilson
quark flavours, tuned to be degenerate in the continuum.
Comparing pseudoscalar masses, obtained from connected quark diagrams made of
tmQCD and/or standard Wilson quark propagators, we confirm that flavour symmetry breaking
effects, which are at most $5\%$, decrease as we approach the 
continuum limit. We also compute the pseudoscalar decay constant in 
the continuum limit, with reduced systematics. As a consequence 
of improved tuning of the mass parameters at $\beta = 6.1$, we reanalyse 
our previous $B_K$ results. Our main phenomenological findings are
$r_0 f_K~=~0.421(7)$ and $\hat B_K = 0.735(71)$.}
\vskip 3 true mm
\thicktablerule
\vspace{10truemm}
\eject
\end{titlepage}

\section{Introduction}
\label{sec:intro}

$B_K$, the bag parameter of neutral $K$-meson oscillations, has been 
computed with several discretizations of lattice fermions. Until recently,
the quenched Wilson fermion results of $B_K$ were the least accurate, 
due to a limited control of those systematic sources of error, which arise 
form the lack of chiral symmetry in the regularization. This trend
has been reversed in ref.~\cite{Dimopoulos:2006dm}, thanks to the
implementation of twisted Wilson fermions~\cite{tmqcd:pap1}.
The simulations of ref.~\cite{Dimopoulos:2006dm}, besides 
introducing some novelties in the computation of $B_K$
(twisted mass QCD (tmQCD) regularization, Schr\"odinger functional
renormalization and RG running) have also been extensive 
from the computational point of view. In particular two tmQCD variants
of the fermion action have been implemented (with twist angles $\pi/2$
and $\pi/4$) at several inverse gauge couplings $\beta$. Such a large
collection of data enables us to address, in the present work, several other
issues related to the tmQCD formalism, in the region of strange quarks. 
Clearly, the fact that we work in the quenched approximation is a limitation
of the scope of the present work.
\parbreak
In sect.~\ref{sec:tmQCD-gen} we present the details of our formalism.
In our first tmQCD variant we introduce two twisted flavours, with
twist angle $\pi/2$, and two standard (untwisted) flavours. In the
second variant we only have two twisted flavours with twist angle
$\pi/4$. The tmQCD lattice action is $O(a)$ improved. All dimension-3
operators (currents, scalar and pseudoscalar densities) are improved
by introducing $O(a)$ Symanzik counterterms. This is essential to
$O(a)$ improvement in the $\pi/4$ case and to those quantities of the
$\pi/2$ case which are not exclusively composed of fully twisted
quarks. No improvement of the four-fermion operators is attempted.
\parbreak
In sect.~\ref{sec:twangle}, we first of all examine the quality of the 
tuning of the mass parameters, fixed so as to ensure that the twist 
angle is equal to a target reference value (in our case $\pi/2$ or 
$\pi/4$). This, together with the requirement that all quark masses be 
degenerate and phenomenological quantities
be measured at a reference pseudoscalar meson mass, constitute our
constant physics requirement. It has to be maintained as we increase
the theory's UV cutoff (i.e. approach the continuum limit).
The twist angle is then measured with the aid of Ward identities and
compared to its target reference value, used in the mass parameter
tuning. We see that $O(a^2)$ cutoff effects are responsible for
statistically significant discrepancies, which are nevertheless
only a few percent. 
\parbreak
In sect.~\ref{sec:flavbreak}, we measure the pseudoscalar masses and decay 
constants in both $\pi/2$ and $\pi/4$ setups. In the former case we are also able to
monitor flavour breaking effects, arising from the twisted mass term in the
lattice action. This we do by comparing the pseudoscalar masses and
decay constants for mesons composed exclusively of twisted or 
untwisted valence quarks, as well as those made of one twisted and
one untwisted quark. For the pseudoscalar masses, we find that 
statistically significant effects, which are nevertheless only a few 
percent at the coarsest lattices, disappear as the lattice spacing is 
decreased. Thus flavour symmetry appears to be restored in the continuum limit. 
\parbreak
Our best results for the $K$-meson decay constant are based on a tmQCD
Ward 
identity. They are free of the usual systematic uncertainties arising
from current normalizations and improvement (i.e. ambiguities in the 
values of $\ZA$, $\ZV$ and $\icA$). Extrapolating $\pi/2$ and $\pi/4$
results to a common continuum limit gives a result in full agreement
with earlier estimates. 
\parbreak
The more detailed analysis concerning the accuracy of the tuning of quark 
masses and twist angles, presented here, was performed
after the publication of ref.~\cite{Dimopoulos:2006dm}. 
The quality of the tuning was found to be satisfactory
in all cases, save for the simulations at $\beta=6.1$, mainly in the 
$\pi/4$ case. This is signalled e.g. by relatively large differences
between the value of the target twist angle, set to $\pi/2$ or $\pi/4$
through the tuning of the bare (subtracted) mass parameters, and the value
 obtained by computing the twist angle with the PCAC 
quark mass instead of the subtracted quark mass. The reason for this 
behaviour has been traced back to the value of $\kappa_{\rm cr}$ taken as 
input from the literature. Indeed, for an accurate determination of $\kappa_{\rm cr}$ it
is crucial to fix its $O(a^2)$ ambiguities by following a constant physics condition in the 
approach to the continuum limit. Instead, the value of $\kappa_{\rm cr}(\beta = 6.1)$ 
quoted in \cite{mbar:charm1} comes from an interpolation
of data obtained from a constant physics condition at other values of $\beta$.
While the effect of relaxing the constant physics requirement was found to
be negligible for the data of \cite{mbar:charm1}, its impact on the tuning
of twist angles is large. This is discussed in detail in Appendix \ref{sec:appkcrit}.
\parbreak
The $\beta=6.1$ critical point has been hence determined afresh, 
and $\beta=6.1$ simulations with new mass parameters have been
performed. Thus all results in the present work are generated from the
datasets of ref.~\cite{Dimopoulos:2006dm}, except for those at
$\beta=6.1$, which are completely new. The new $B_K(\beta=6.1)$
results can be found in sect.~\ref{sec:4ferm}. They induce a
reanalysis of the continuum limit extrapolation of this quantity.
Also in sect.~\ref{sec:4ferm}, we collect our detailed results
of the kaon-to-pion four-fermion operator matrix elements, involved
in the $\Delta I =1/2$ rule. Strictly speaking, these results are not
physical, as they refer to four degenerate quarks, with masses close
or above half the strange quark mass. They have been used,
however, in ref.~\cite{Dimopoulos:2006ma} in order to obtain the 
relevant four-fermion operator renormalization with Neuberger
fermions, through a matching procedure of RGI matrix elements
computed from both tmQCD and Neuberger regularizations.
\parbreak
For the sake of legibility, all tables containing our results
have been gathered in Appendix \ref{sec:apptables}.

\section{General tmQCD formalism}
\label{sec:tmQCD-gen}

Twisted mass QCD has been designed to eliminate
exceptional configurations in (partially) quenched 
lattice simulations with light Wilson quarks~\rep{tmqcd:pap1}.
In its original formulation, it describes a
mass-degenerate isospin doublet $\psi$ of Wilson quarks for which, 
besides the standard mass term, a so-called twisted mass term
$i\mu_{\rm q}\bar{\psi}\gamma_5\tau^3\psi$ is introduced. 
The properties of tmQCD have been studied in
detail in~\rep{tmqcd:pap1}, where, in particular, 
its equivalence to standard two-flavour QCD has been established%
\footnote{for reviews on the subject see \cite{Shindler:2005vj,Pena:2006tw,Sint:2007ug}}.
We discuss here the main characteristics of this formulation,
extended to more flavours, in ways analogous to those 
discussed in~\rep{tmqcd:DIrule} and~\rep{Dimopoulos:2006dm}.
\parbreak
It is convenient to formalize our variant of tmQCD in
terms of a twisted and an untwisted isospin doublet, denoted
as $\bar \psi_{\rm tw} = (\bar \psi_1, \bar \psi_2)$ and 
$\bar \psi_{\rm w} = (\bar \psi_3, \bar \psi_4)$ respectively. All flavours
will eventually be tuned to be degenerate.
The twisted (untwisted) isospin doublet is regularized in the standard
tmQCD-Wilson (plain Wilson) fashion:
\begin{align}
 \label{tmQCD_action2}
 S_F = a^4 \sum_x \,\, [ & \ {\bar \psi_{\rm tw}}(x) (D_{\rm w}
+ m_{0,{\rm tw}} + i\gamma_5 \tau^3 \mu_{0,{\rm tw}} )\psi_{\rm tw}(x) \nonumber  \\
& + \bar \psi_{\rm w}(x) (D_{\rm w} + m_{0,{\rm w}} ) \psi_{\rm w}(x)\ ] \ ,
\end{align}
where $D_{\rm w}$ is the standard Dirac-Wilson fermion matrix (with
a Clover term) and $\tau^3$ the Pauli isospin matrix.
In this work, the Wilson plaquette action is the regularization
of the pure gauge sector of the theory. For the rest of the notation,
relating tmQCD to standard QCD (concerning field rotations,
mass transformations etc.) see~\rep{Dimopoulos:2006dm}.
Here, what we are mostly interested in are the expressions for
the renormalized quark mass in the twisted quark sector,
given by the combination of standard and twisted mass parameters
\begin{eqnarray}
M_{\rm R,tw} = \sqrt{m_{\rm R,tw}^2 + \mu_{\rm R,tw}^2}\ ,
 \label{quark_mass_M}
\end{eqnarray}
and the twist angle, defined in terms of renormalized masses
as
\begin{equation}
 \tan \taa = \dfrac{\muren{,{\rm tw}}}{\mren{,{\rm tw}}} \,\, .
 \label{eq:tan}
\end{equation}
\parbreak
Also standard is the relation between renormalized and bare
quark masses: the subtracted (unrenormalized) quark mass for 
Wilson fermions in denoted by 
$a m_{\rm q} = 1/(2 \kappa) - 1/(2 \kappa_{\rm cr})$, $\kappa$ 
being the hopping parameter ($2\kappa = [a m_0 + 4]^{-1}$).
Whenever we need to identify the
quark doublet $f$ (with $f = {\rm tw},{\rm w}$), we will denote the corresponding quantities by
$a m_{{\rm q},f}$ and $\kappa_f$.  In a Symanzik $O(a)$ improved
framework, the renormalized quenched quark masses 
for the untwisted flavours are given by
\begin{equation}
\mren{,{\rm w}} = \Zm[ m_{{\rm q},{\rm w}} ( 1 + \ibm a m_{{\rm q},{\rm w}} ) ]\ ,
\label{eq:msren}
\end{equation}
while the twisted quark masses renormalize as follows:
\begin{eqnarray}
\mren{,{\rm tw}} &=& \Zm [ m_{{\rm q},{\rm tw}} ( 1 + \ibm a m_{{\rm q},{\rm tw}} ) + {\tilde b}_{\rm m} a \mu_{0,{\rm tw}}^2]\ , 
\label{eq:mlrenu}
\\
\muren{,{\rm tw}} &=& \ZP^{-1} \mu_{0,{\rm tw}} ( 1 + b_\mu a m_{{\rm q},{\rm tw}} )\ .
\label{eq:mlren}
\end{eqnarray}
In terms of the last two expressions, the twist angle may be expressed
as~\cite{tmqcd:pap2}
\begin{eqnarray}
\tan(\alpha) = \dfrac{a \mu_{0,{\rm tw}} [1 + (b_\mu - \ibm) a m_{{\rm q},{\rm tw}}]}
{\ZP\Zm [a m_{{\rm q},{\rm tw}} + {\tilde b}_{\rm m} (a \mu_{0,{\rm tw}})^2 ]} \,\, .
\label{eq:tana}  
\end{eqnarray}
\parbreak
We now discuss an alternative expression for the twist angle, obtained
in terms of Ward identities in~\rep{tmqcd:pap2}.
For the twisted sector of our theory, the PCAC and PCVC Ward identities read
\begin{eqnarray}
\tilde \partial_\nu (A_{\rm R})_{\nu,12} &=& 2 m_{\rm R} (P_{\rm R})_{12}\ ,
\label{eq:pcac}
\\
\tilde \partial_\nu (V_{\rm R})_{\nu,12} &=& 2 i \mu_{\rm R} (P_{\rm R})_{12}\ ,
\label{eq:pcvc}
\end{eqnarray}
and they are valid up to $O(a^2)$ for the Symanzik-improved
operators~\cite{tmqcd:pap2}
\begin{eqnarray}
(P_{\rm R})_{12} &=& \ZP [1 + \ibP a m_{{\rm q},{\rm tw}} ] P_{12}\ ,
\label{eq:Prenimp} \\
(A_{\rm R})_{\nu,12} &=& \ZA [1 + \ibA a m_{{\rm q},{\rm tw}} ] [ A_{\nu,12}
+ a \icA \tilde \partial_\nu P_{12} - i {\tilde b}_{\rm\scriptscriptstyle A} a \mu_{0,{\rm tw}} V_{\nu,12}]\ ,
\label{eq:Arenimp} \\
(V_{\rm R})_{\nu,12} &=& \ZV [1 + \ibV a m_{{\rm q},{\rm tw}} ] [ V_{\nu,12}
+ a \icV \tilde \partial_\rho T_{\nu\rho,12} - i {\tilde b}_{\rm\scriptscriptstyle V} a \mu_{0,{\rm tw}} A_{\nu,12}]\ ,
\label{eq:Vrenimp}
\end{eqnarray}
where $\tilde \partial_\nu$ denotes the lattice symmetrized
derivative and the operator subscripts $1,2$ indicate quark flavours. 
By combining expressions~(\ref{eq:pcac}) - (\ref{eq:Vrenimp}) with the 
standard definition of the PCAC bare quark mass~\cite{impr:pap1}, 
\begin{eqnarray}
\tilde \partial_\nu [A_{\nu,12} + \icA a 
\tilde \partial_\nu P_{12} ] = 2 m_{\rm tw} P_{12} \ ,
\label{eq:mpcac}
\end{eqnarray}
the renormalized quark mass is computed as\footnote{
\req{eq:mpcac} is to be understood 
in terms of correlation functions involving these
operator insertions. In the present work Schr\"odinger
functional correlation functions with pseudoscalar
boundaries are implemented. The same considerations hold for
\req{eq:mpcvc}
}
\begin{eqnarray}
\mren{,{\rm tw}} = \dfrac{\ZA[1 + \ibA am_{{\rm q},{\rm tw}}]}{\ZP [1+ \ibP am_{{\rm q},{\rm tw}}]}
\Big [ m_{\rm tw} + a \mu_{0,{\rm tw}}^2 {\tilde b}_{\rm\scriptscriptstyle A} \ZV^{-1} \Big ]\ .
\label{eq:mrenpcac}
\end{eqnarray}
Finally, the PCAC expression for the twist angle, in terms
of the above, is~\cite{tmqcd:pap2}
\begin{eqnarray}
\tan(\alpha) = \dfrac{a \mu_{0,{\rm tw}} [1 + (b_\mu + \ibP - \ibA) a m_{{\rm q},{\rm tw}}]}
{\ZA [a m_{\rm tw} + \tilde b_{\rm\scriptscriptstyle A} (a \mu_{0,{\rm tw}})^2 \ZV^{-1}]} \,\, .
\label{eq:tanapcac}  
\end{eqnarray}
\parbreak
Yet another variant of the twist angle is obtained by expressing
$\muren{,{\rm tw}}$ in terms of the PCVC Ward identity. We define the PCVC
bare twisted mass from\footnote{The tensor term proportional to $\icV$
in~\req{eq:Vrenimp} vanishes upon differentiation of the
vector current $(V_{\rm R})_{\nu,12}$.}
\begin{eqnarray}
\tilde \partial_\nu V_{\nu,12} = 2i\mu_{\rm tw} P_{12} \ ,
\label{eq:mpcvc}
\end{eqnarray}
and, using expressions~(\ref{eq:pcac}) - (\ref{eq:Vrenimp}),
obtain for the renormalized twisted mass
\begin{eqnarray}
\muren{,{\rm tw}} = \dfrac{\ZV [1 + \ibV am_{{\rm q},{\rm tw}}]}{\ZP [1+ \ibP am_{{\rm q},{\rm tw}}]}
\Big [ 1 - a m_{\rm tw} \ZV {\tilde b}_{\rm\scriptscriptstyle V} \Big ]\mu_{\rm tw}\ .
\label{eq:mrenpcvc}
\end{eqnarray}
The PCVC expression for the twist angle, in terms
of the above, is
\begin{eqnarray}
\tan(\alpha) = \dfrac{\ZV a \mu_{\rm tw} [1 + (\ibV - \ibA - \ZV\tilde b_{\rm\scriptscriptstyle V}) a m_{{\rm q},{\rm tw}}]}
{\ZA [a m_{\rm tw} + \tilde b_{\rm\scriptscriptstyle A} (a \mu_{\rm tw})^2 \ZV^{-1} ]} \,\, .
\label{eq:tanapcvc}  
\end{eqnarray}
\parbreak
In the present work, expression~(\ref{eq:tana}) is used for
tuning the bare mass parameters $\mu_{0,{\rm tw}}$ and $\kappa_{\rm tw}$,
so as to fix the theory to a specific twist angle. One also needs
the value of $\kappa_{\rm cr}$ for the determination of $a m_{{\rm q},l}$.
This is known from previous Schr\"odinger functional computations,
based on the Clover-improved theory with Wilson (untwisted)
quarks.\footnote{The condition we implement
for fixing the twist angle is neither the so-called pion mass determination,
nor the PCAC one of ref.~\cite{Jansen:2005gf}.}
Once the bare parameters are thus
fixed to satisfy~\req{eq:tana}, one may use~\req{eq:tanapcac} 
and~\req{eq:tanapcvc} in order to obtain independent estimates
of the twist angle, which differ from the target value by
$O(a^2)$ discretization effects. This provides a measure of the
systematic uncertainties related to the tuning of the twist
angle.\footnote{The (re)normalization
constants ($\ZA$, $\ZP$ etc.) as well as the improvement coefficients
($\icA$, $\ibA$ etc.) used in this work are also taken
from previous Schr\"odinger functional computations. Their values
are all gathered in Appendix~A of ref.~\cite{Dimopoulos:2006dm}.
Note that~Eqs.~(A.10) and (A.11) of that Appendix contain
misprints; they should
read $\tilde b_{\rm\scriptscriptstyle A} = 0.086 C_F g_0^2$ and  
$\tilde b_{\rm\scriptscriptstyle V} = 0.074 C_F g_0^2$.}
\parbreak
Finally, we specify the twist angles, following the two cases
of~\rep{Dimopoulos:2006dm}. In the first case, known as
fully twisted theory, the bare parameters
of the twisted quark doublet are tuned so as to ensure that $\alpha =\pi/2$.
This amounts to tuning Eq.~(\ref{eq:mlrenu}) so that $\mren{,{\rm tw}} = 0$.
The untwisted doublet is also tuned (through an opportune
choice of $\kappa_{\rm w}$), so that $\mren{,{\rm w}} = \muren{,{\rm tw}}$
up to $O(a^2)$; cf. Eqs.(\ref{eq:msren}) and (\ref{eq:mlren}).
In the second case we switch off the untwisted doublet, keeping
just two twisted flavours; the twist
angle is set to $\alpha = \pi/4$. This amounts to tuning
Eqs.(\ref{eq:mlrenu}) and (\ref{eq:mlren}) so as to have
$\mren{,{\rm tw}} = \muren{,{\rm tw}}$. The detailed expressions
used for these tunings may be read off from
ref.~\rep{Dimopoulos:2006dm}.
\parbreak
In order to make contact with the physical results we will be presenting,
we now identify the four generic quark fields $\psi_k$ ($k= 1,\cdots,4$)
with physical (if degenerate) flavours. There is a different
identification according to the problem in hand. When we discuss
pseudoscalar masses, decay constants and $B_K$,
in the $\pi/2$ case we identify the twisted
doublet with $\bar \psi_{\rm tw} = ( \bar u , \bar d)$ and the
untwisted one with $\bar \psi_{\rm w} = ( \bar s , \bar c)$.
The same quantities in the $\pi/4$ case are addressed in terms of a
single twisted doublet of a strange and a down
quark; i.e. $\bar \psi_{\rm tw} = (\bar s , \bar d)$.
As our simulations are quenched and mass degenerate, this
is sufficient to model two valence quarks. In this way we
make full contact with the notation of the earlier tmQCD
simulations of ref.~\cite{Dimopoulos:2006dm}.
For the results related to $\Delta S = 1$ four-fermion operators,
the physical flavour identification is more complicated.
The reader is referred directly to ref.~\cite{Dimopoulos:2006ma},
where the issue has been addressed in detail.

\section{Cutoff effects of the twist angle}
\label{sec:twangle}

\begin{figure}[ht]
\begin{center}
\epsfig{figure=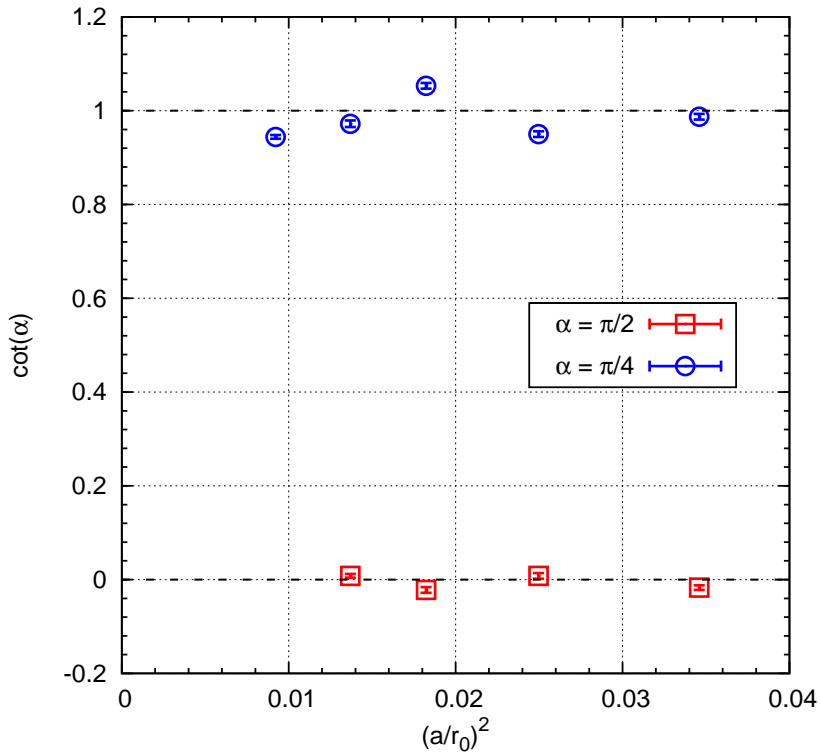, angle=-90, width=11.0 true cm}
\caption{Lattice spacing dependence of $\cot(\taa)$, computed at the
  physical kaon mass point, for twist angles $\pi/2$ (circles)
  and $\pi/4$ (squares). The errors are smaller than the points.}
\label{fig:cot}
\end{center}
\end{figure}

We now turn to the determination of the twist angle from PCAC and PCVC
relations. The simulation parameters are gathered in
Table~\ref{tab:runspi2} for the $\pi/2$ theory and in 
Table~\ref{tab:runspi4} for the $\pi/4$ one.
The data are those of ref.~\cite{Dimopoulos:2006dm},
except at $\beta = 6.1$. As mentioned in the introduction, the run
had to be repeated at this coupling, for reasons which will be discussed 
in detail below. The physical regime targeted in the runs of
ref.~\cite{Dimopoulos:2006dm} is that of the $K$-meson, composed
of two degenerate valence quarks. As detailed in that work,
in practice this means that the choice of quark masses is such 
that the $K$-meson in the $\pi/2$ theory
is computed in the range $640$--$830~\MeV$ and extrapolated
to the physical point at 495 MeV (i.e. $r_0 M^{\rm phys}_K = 1.2544$).
In the $\pi/4$ theory we are instead able to simulate with quarks
corresponding to a physical kaon of about 495 MeV. The only
exception is the $\beta = 6.45$ case, in which extrapolations
from higher mass values were the only option, as simulations 
with quarks corresponding to a physical kaon require prohibitively 
large lattice sizes. As the present work is based on the same runs, 
the same physical point for all four flavours is targeted here.
\parbreak
Following the discussion of sect.~\ref{sec:tmQCD-gen},
the bare mass parameters (i.e. standard 
hopping parameter(s) and twisted mass $\mu_{0,{\rm tw}}$) are tuned 
at each $\beta$ value
so as to keep the quarks degenerate and the
twist angle fixed at $\taa = \pi/2, \pi/4$.
The $\kappa_{\rm cr}$ estimates used in the bare parameter
calibration are taken over from the literature; see
Table~\ref{tab:kcrit}. Those of ref.~\cite{impr:pap3},
as well as the one provided to us by the ZeRo 
Collaboration\footnote{We thank I.~Wetzorke for
providing us with the  $\kappa_{\rm cr}(\beta = 6.3)$ value,
obtained in the context of~ref\cite{Guagnelli:2004ga}.}, are
the result of a direct computation of the PCAC Ward identity at the
corresponding bare coupling, using $O(a)$ improved (untwisted) Wilson
fermions, whereas the estimates from
ref.~\cite{mbar:charm1} are the result of an interpolation in $\beta$
of the data of ref.~\cite{impr:pap3}. We will see that this
$\kappa_{\rm cr}$ estimate at $\beta = 6.1$ results in poor
tuning of the bare tmQCD parameters.
Thus, following the procedure described in ref.~\cite{impr:pap3}, we
have recomputed $\kappa_{\rm cr}$ at $\beta = 6.1$, finding a
different value. This has been done by an independent run, performed 
at lattice volume $L/a=16$, $T/L=2$ and four PCAC quark mass values in a range 
$0.025 \lesssim am_{\rm av} \lesssim 0.05$, similar to that of ref.~\cite{impr:pap3}
(where $m_{\rm av}$ is defined). Our $\kappa_{\rm cr}$ result is obtained 
by linear exrapolation in the PCAC quark masses, ensuring that 
our systematics resemble those of ref.~\cite{impr:pap3}. 
\parbreak
All observables computed in this work are obtained from the large
time asymptotic limit of operator correlation functions with
Schr\"odinger functional (SF) boundary conditions.
The notation is standard, following closely that adopted
in  e.g.~\rep{Dimopoulos:2006dm}. For instance, $f_{\scrA,12}$ denotes
the Schr\"odinger functional correlation with a fermionic operator
$A_{0,12}$ in the bulk and a time-boundary pseudoscalar
operator $\bar \zeta_2 \gamma_5 \zeta_1$ at $x=0$.
All such correlation functions are properly (anti)symmetrized in time,
when used to extract quark masses, effective pseudoscalar masses
and decay constants. The bare PCAC and PCVC quark masses of~Eqs.~(\ref{eq:mpcac})
and~(\ref{eq:mpcvc}) are obtained from the ratios
\begin{eqnarray}
R_{m_{\rm tw}} &=& \dfrac{\tilde \partial_0 [f_{\scrA,12}(x_0) + \icA a
\tilde \partial_0 f_{\scrP,12}(x_0)]}{2  f_{\scrP,12}(x_0)}\ , \\
R_{\mu_{\rm tw}} &=& -i \dfrac{\tilde \partial_0 f_{\scrV,12}(x_0)}{2
  f_{\scrP,12}(x_0)} \,\, .
\end{eqnarray}
From these and Eqs.~(\ref{eq:mrenpcac}), (\ref{eq:mrenpcvc}) 
we compute the renormalized masses $a \mren{,{\rm tw}}$ and $a \muren{,{\rm tw}}$; from
Eqs.~(\ref{eq:tanapcac}), (\ref{eq:tanapcvc}) we compute
$\cot(\alpha)$. These results are reported in Table~\ref{tab:cota}.
The errors are statistical; we have checked that systematic errors
due to the uncertainties of $\kappa_{\rm cr}$, the (re)normalization
parameters and the improvement coefficients are an order of magnitude
smaller than statistical ones.
\parbreak
Let us comment on our $\pi/2$ results first. The two $a \muren{,{\rm tw}}$
values (computed from \req{eq:mlren} and \req{eq:mrenpcvc}) have
a small but statistically significant discrepancy of less than 3\%.
Similarly, the result for $a \mren{,{\rm tw}}$ is also statistically
different from the target value $a \mren{,{\rm tw}} = 0$.
The two estimates of the twist angle are only a few percent off the
target  value $\pi/2$, but the small discrepancies in the 
$a \muren{,{\rm tw}}$ evaluations do not carry over to the two results for 
$\cot(\taa)$, which are nearly always compatible within errors.
All observed deviations from the expected target values
are attributed to $O(a^2)$ discretization effects. 
\parbreak
The $\pi/4$ results display largely the same characteristics,
but the mass range spanned by the data (for $\beta < 6.45$) is too narrow
to discern the details of the  dependence of the twist angle upon the quark
mass parameters. What we see is that the difference between the two twisted mass 
estimates $\muren{,{\rm tw}}$ and the standard mass estimate 
$\mren{,{\rm tw}}$  is a small but statistically significant effect. 
The same is true of the two Ward identity estimates of the twist angle,
which also differ from the target value $\pi/4$ by a small amount.
\parbreak
Finally, we study to which extent the Ward identity estimates
of the twist angles approach their target values $\pi/2, \pi/4$
in the continuum limit. We do this by first computing,
at fixed $\beta$ value, the quantity $\cot(\taa)$ at the kaon mass
reference scale $M_K \equiv M^{\rm phys}_{K} = 495$~MeV. 
This is done by extra/interpolating
our data as a function of the pseudoscalar effective mass-squared,
expressed in physical units.\footnote{The pseudoscalar mass in
question is obtained from a correlation function consisting exclusively
of twisted quark propagators. In the next section, this effective
mass is denoted as $(r_0 M^{\rm eff}_{ud})$ for the $\pi/2$
case, while for the $\pi/4$ one we use $(r_0 M^{\rm eff}_{sd})$.}
The result thus obtained is plotted against $\left(a/r_0\right)^2$ in 
Fig.~\ref{fig:cot}. A linear extrapolation of $\cot(\taa)$
to the continuum limit
turns out to be unreliable (very large $\chi^2/{\rm d.o.f.}$),
as the data do not display a monotonic behaviour, with
fluctuations which are much larger than their errors. 
In any case, these fluctuations are only a small effect,
reflecting the overall uncertainty of
the tuning of the twist angle to a constant target value
(which amounts to a condition of constant physics as we approach
the continuum limit).

\section{Flavour symmetry breaking, meson masses and decay constants}
\label{sec:flavbreak}

\begin{figure}
\hskip -1.5cm\epsfig{figure=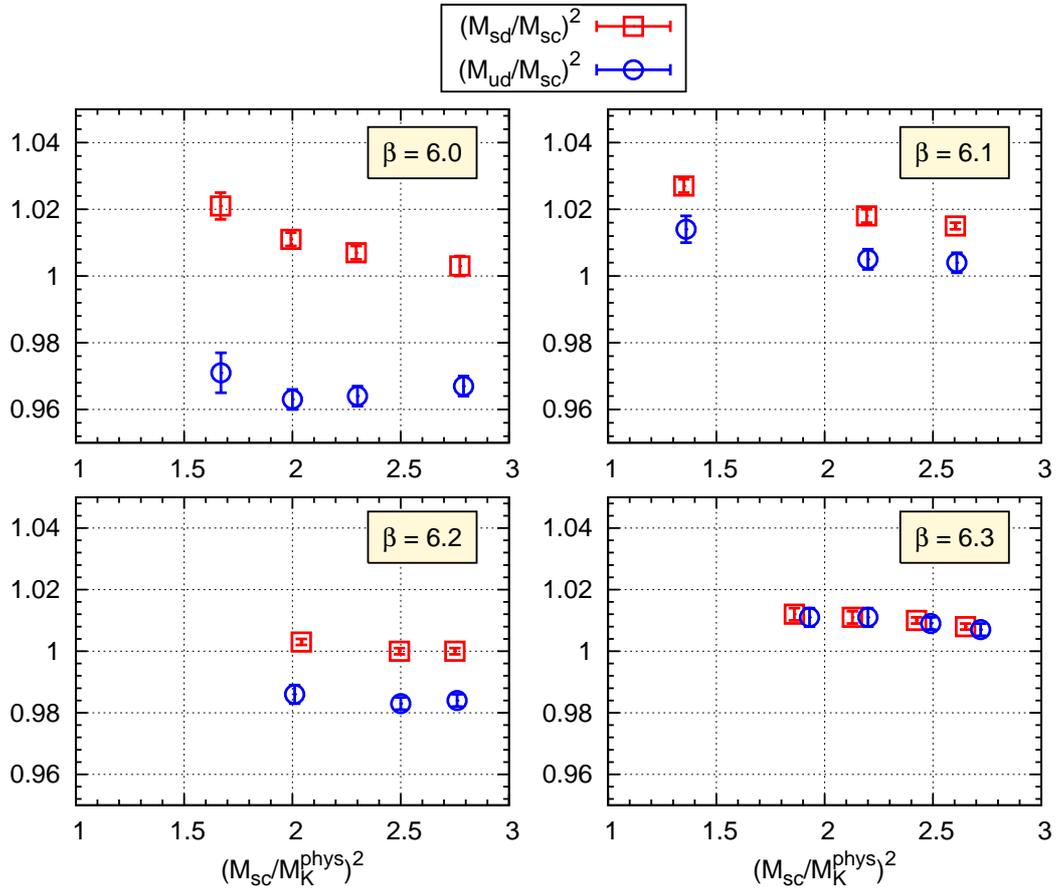, angle=-90, width=17.5 true cm}
\begin{center}
\caption{Flavour breaking effects for the quantities
  $\left(M_{sd}/M_{sc}\right)^2$ (squares) and $\left(M_{ud}/M_{sc}\right)^2$ (circles), as
  functions of $\left(M_{sc}/M^{\rm phys}_{K}\right)^2$, in the $\pi/2$ theory. In the
  $\beta=6.3$ plot, the abscissae of the circles are slightly displaced for clarity.}
\label{fig:isomass}
\end{center}
\end{figure}

In order to investigate flavour breaking effects,
we compare effective pseudoscalar masses,
obtained from correlation functions with different 
combinations of twisted and standard Wilson quark propagators,
on our quenched configuration ensemble of tmQCD at $\taa=\pi/2$.
The rationale behind these computations is as follows:
we are dealing with a (quenched) lattice QCD model with
four degenerate quark flavours, two of which are twisted.
The massless continuum theory has an $SU(4)_{\rm L} \otimes SU(4)_{\rm R}$
chiral symmetry with 15 degenerate pseudoscalar Goldstone bosons;
with massive fermions we are left with the vector flavour symmetry $SU(4)_{\rm V}$.
In the regularized theory, the Wilson term breaks the symmetry
induced by axial transformations, even in the absence of quark mass parameters.
In twisted mass QCD some of these axial symmetries are interpreted as
part of the SU(4) flavour symmetry. With our choice of twisted mass terms one 
then finds that on the lattice the SU(4) flavour symmetry is reduced 
to the subgroup  $U(1)\otimes U(1) \otimes SU(2)$. Thus, at finite lattice 
spacing, we expect that
the ``twisted'' charged Goldstone bosons, the ``untwisted'' ones and
the ``twisted-untwisted'' ones will differ in mass by terms
which are like $O(a^2)$ (recall that we work with
a Symanzik-improved action). Based on this approach, we have
provided a first summary of our findings on this symmetry breaking
in~\rep{Dimopoulos:2006dm}; here we present our full results.
The same approach for monitoring such flavour breaking effects
has also been implemented (with an action with two $\pi/2$-twisted
isospin doublets and no Clover term) in~\rep{Abdel-Rehim:2006ve}.
Both works focus on pseudoscalar masses in the kaon region.
For similar results closer to the chiral limit (for a single
$\beta$ value), see~\rep{Becirevic:2006ii}.
It is fairly straightforward to measure these flavour breaking effects, 
as the corresponding correlation functions involve only connected diagrams. 
Recall that there is also a flavour breaking effect between the 
charged and neutral ``twisted'' pseudoscalar, which is harder to 
monitor, as the measurement of the neutral pion mass involves 
disconnected diagrams; see \cite{Farchioni:2005hf} Studying this 
flavour breaking is beyond the scope of the present work.
\parbreak
We have measured pseudoscalar
effective masses $a M^{\rm eff}_{ij} (x_0)$, 
using suitable time-dependent correlation functions $f_{(\scrA_{\rm R})_{ij}}(x_0)$ 
(with $i,j$ distinct flavour indices) of the temporal component of
the renormalized axial current $(A_{\rm R})_{ij}$
\begin{equation}
a M^{\rm eff}_{ij} (x_0) = \dfrac{1}{2} 
\ln \Bigg [ \dfrac{ f_{(\scrA_{\rm R})_{ij}} (x_0-a)}
{f_{(\scrA_{\rm R})_{ij}} (x_0+a)} \Bigg ] \,\, .
\label{eq:effmass}
\end{equation}
These quantities are suitably (anti)symmetrized in time
and averaged over plateaux, as detailed in~\rep{Dimopoulos:2006dm}.
In the language of the twisted action~\req{tmQCD_action2}
at twist angle $\pi/2$, the corresponding lattice correlation
functions to be used in \req{eq:effmass} are:
\begin{itemize}
\item $f_{(\scrA_{\rm R})_{sd}}(x_0) \rightarrow
\dfrac{1}{\sqrt 2}[f_{(\scrA_{\rm R})_{sd}} (x_0) -i f_{(\scrV_{\rm R})_{sd}} (x_0)]$ \ ,
\item $f_{(\scrA_{\rm R})_{ud}}(x_0) \rightarrow - i f_{(\scrV_{\rm R})_{ud}} (x_0)$\ ,\\[-8pt]
\item $f_{(\scrA_{\rm R})_{sc}}(x_0) \rightarrow f_{(\scrA_{\rm R})_{sc}} (x_0)$\ .
\end{itemize}
The first correlation function is composed of a tmQCD Wilson
quark propagator and a standard QCD Wilson quark propagator. 
From it we derive the ``$K$-meson'' effective mass, denoted by 
$a M^{\rm eff}_{sd}$. The second correlation function, composed 
exclusively of the tmQCD Wilson quark propagator, provides the 
charged ``pion'' effective mass $a M^{\rm eff}_{ud}$. Finally,
the third correlation function, composed exclusively of the
standard QCD Wilson quark propagator, provides the ``$D_s$-meson''
effective mass $a M^{\rm eff}_{sc}$. Since all quark masses are
tuned to be degenerate, these are three of the 15 degenerate Goldstone
bosons in the continuum limit of our quenched theory. At finite lattice
spacing tmQCD induces flavour breaking discretization effects, which are
monitored by comparing the values of the three effective masses.
\parbreak
The corresponding currents, inserted in the above correlation functions,
are the following Symanzik-improved
quantities, taken over from Appendix B of~\rep{Dimopoulos:2006dm}:
\begin{eqnarray}
(A_{\rm R})_{\nu,sd} &=& \ZA [1 + \dfrac{1}{2} \ibA a m_{{\rm q},{\rm w}} ] [ A_{\nu,sd}
+ a \icA \tilde \partial_\nu P_{sd} - i \dfrac{1}{2} \tilde b_{\rm\scriptscriptstyle A} a \mu_{0,{\rm tw}} V_{\nu,sd}]\ ,
\\
(V_{\rm R})_{\nu,sd} &=& \ZV [1 + \dfrac{1}{2} \ibV a m_{{\rm q},{\rm w}} ] [ V_{\nu,sd}
- i \dfrac{1}{2} \tilde b_{\rm\scriptscriptstyle V} a \mu_{0,{\rm tw}} A_{\nu,sd}]\ ,
\\[1pt]
(V_{\rm R})_{\nu,ud} &=& \ZV [ V_{\nu,ud} - i \tilde b_{\rm\scriptscriptstyle V} a \mu_{0,{\rm tw}} A_{\nu,ud}]\ ,
\\[6pt]
(A_{\rm R})_{\nu,sc} &=& \ZA [1 + \ibA a m_{{\rm q},{\rm w}} ] [ A_{\nu,sc}
+ a \icA \tilde \partial_\nu P_{sc}]\ .
\label{eq:axsc}
\end{eqnarray}
Again the $O(a)$ tensor-like counterterm in the above vector currents has
been omitted, since it drops out in the correlation functions (it is a sum
over space, with periodic boundary conditions, of a discrete spatial divergence).
\parbreak
Another estimate of the effective pseudoscalar meson mass, denoted as
$a \tilde M^{\rm eff}_{ud}$, is obtained by using the correlation 
$f_{(\scrP_{\rm R})_{ud}}$ in \req{eq:effmass}, with
\begin{equation}
(P_{\rm R})_{ud} = \ZP [ 1 + \ibP a m_{{\rm q},{\rm tw}} ] P_{ud}\ .
\end{equation}
\parbreak
The results are collected in Table~\ref{tab:meffpi/2}. 
The agreement between $M^{\rm eff}_{ud}$ and 
$\tilde M^{\rm eff}_{ud}$ is excellent.
In Fig.~\ref{fig:isomass} the ratio $\left(M_{sd}/M_{sc}\right)^2$,
plotted against $\left(M_{sd}/M^{\rm phys}_{K}\right)^2$ with 
$M^{\rm phys}_{K} = 495$ MeV, is compared to $\left(M_{ud}/M_{sc}\right)^2$.
Note that the statistical errors of these ratios turn out to be very
small, due to the strong correlations between numerator and
denominator. We see that at $\beta = 6.0$ the two ratios are
incompatible; their deviation from unity (which quantifies
flavour breaking) is at most a $5\%$ effect. As we approach
the continuum limit at $\beta = 6.3$, the two ratios
become compatible and their
deviation from unity reduces to $1-2\%$. Our conclusion is that
flavour symmetry breaking effects at the mass ranges we are
considering appear to be under control, diminishing fast as the
continuum limit is approached. Besides this general conclusion,
there are a couple of observations to be made:
(i) At $\beta = 6.0$ we find that $M_{ud} < M_{sd}$, in agreement 
to the findings of~\rep{Abdel-Rehim:2006ve} (see Fig.~13 of that 
work)~\footnote{The authors of~\rep{Abdel-Rehim:2006ve} call
$M_{K^+}$ what we call $M_{ud}$ and ``kaon with Wilson strange''
what we call $M_{sd}$.};
(ii) at fixed reference mass $(M_{sc}/M^{\rm phys}_{K})^2$, 
these mass ratios do not display a monotonic dependence on $\beta$. 
This is probably a small cumulative effect of the many systematic
uncertainties of the mass tuning procedure.
\parbreak
A similar analysis is performed for the pseudoscalar meson decay
constants. In the Schr\"odinger functional framework they are
obtained from the axial current correlation
functions $f_{\scrA_{\rm R}}$, properly normalized by the boundary-to-boundary
correlation function $f_1$; see~\rep{mbar:pap2} for details.
For the tmQCD $\pi/2$ case under investigation, the specific expressions
are (in the large time asymptotic regime):
\begin{eqnarray}
F_{sd} &\approx&
{\sqrt 2} (M^{\rm eff}_{sd} L^3)^{-1/2} \exp[
(x_0 - T/2)M^{\rm eff}_{sd}] \dfrac{f_{(\scrA_{\rm R})sd}(x_0) - i f_{(\scrV_{\rm R})sd}(x_0)}{\sqrt{f_{1,sd}}}\ ,
\label{eq:dc-sd}
\\
F_{ud}  &\approx&
2 (M^{\rm eff}_{ud} L^3)^{-1/2} \exp[
(x_0 - T/2)M^{\rm eff}_{ud}] \dfrac{- i f_{(\scrV_{\rm R})ud}(x_0)}{\sqrt{f_{1,ud}}}\ ,
\label{eq:dc-ud}
\\
F_{sc} &\approx& 2 (M^{\rm eff}_{sc} L^3)^{-1/2} \exp[
(x_0 - T/2)M^{\rm eff}_{sc}] \dfrac{f_{(\scrA_{\rm R})sc}(x_0)}{\sqrt{f_{1,sc}}}\ .
\label{eq:dc-sc}
\end{eqnarray}
The quantities $F_{sd}, F_{ud}$ and $F_{sc}$ are obtained from these
expressions in a range of $x_0 \gg 0$ in which the
pseudoscalar effective masses have been extracted.
\parbreak
A second method for computing $F_{ud}$ is based on the PCVC
relation~\req{eq:pcvc}, expressed in terms of Schr\"odinger
functional correlation functions:
\begin{equation}
-M_{ud} f_{(\scrV_{\rm R})_{ud}} (x_0) = 2 i \muren{,{\rm tw}} f_{(\scrP_{\rm R})_{ud}} (x_0)\ .
\label{eq:exctPCVC2}
\end{equation}
The corresponding decay constant is computed as
\begin{eqnarray}
\tilde F_{ud}  &\approx& - 4
\dfrac{\muren{,{\rm tw}}}{\tilde M^{\rm eff}_{ud}}
(\tilde M^{\rm eff}_{ud} L^3)^{-1/2} \exp[
(x_0 - T/2)\tilde M^{\rm eff}_{ud}] \dfrac{f_{(\scrP_{\rm R})ud}(x_0)}{\sqrt{f_{1,ud}}}\ .
\label{eq:dc-ud2}
\end{eqnarray}

\begin{figure}
\hskip -1.5cm\epsfig{figure=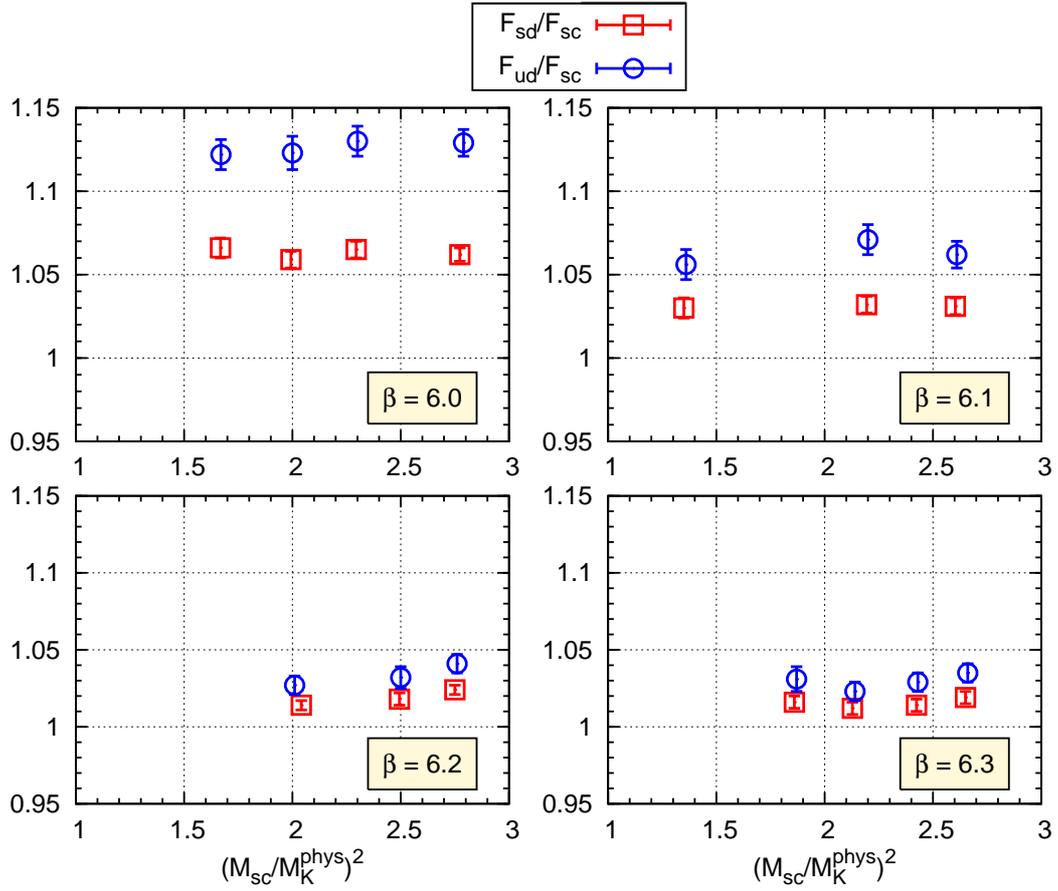, angle=-90, width=17.5 true cm}
\begin{center}
\caption{Flavour breaking effects for the quantities
  $F_{sd}/F_{sc}$ (squares) and $F_{ud}/F_{sc}$ (circles), as
  functions of $\left(M_{sc}/M^{\rm phys}_K\right)^2$, in the $\pi/2$ theory.}
\label{fig:isodecay}
\end{center}
\end{figure}

The results for the decay constants are collected in
Table~\ref{tab:decpi2}. Note the excellent agreement
between the $F_{ud}$ and $\tilde F_{ud}$ results, 
at all $\beta$ values, which is analogous to the
agreement between the two $\cot(\taa)$ computations
presented in sect.~\ref{sec:twangle}.
In Fig.~\ref{fig:isodecay}
the ratio $F_{sd}/F_{sc}$ is compared to $F_{ud}/F_{sc}$.
The situation is qualitatively analogous to that of the mass ratios
presented above, in that flavour breaking effects tend to vanish as the
continuum limit is approached. These effects range form 13\% to 3\% with
increasing $\beta$; however we show below that these estimates
depend heavily on the improvement coefficent of the axial current.

\begin{figure}
\hskip -0.8cm\epsfig{figure=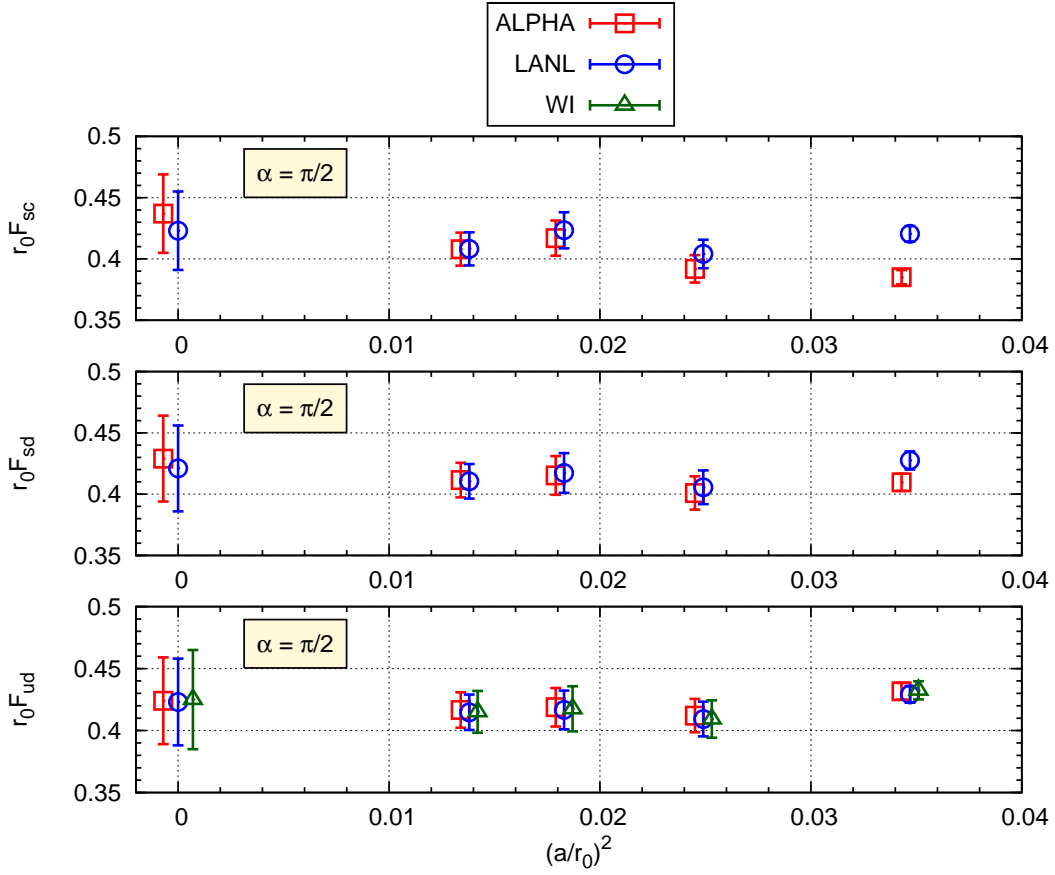, angle=-90, width=16.5 true cm}
\begin{center}
\hskip 2.0cm
\caption{Pseudoscalar meson decay constants, obtained with ALPHA and
  LANL normalization and improvement coefficients at the physical kaon
  mass. The twist angle is $\pi/2$. The results for $r_0 \tilde F_{ud}$ 
  (based on a Ward identity) are also shown. The continuum
  extrapolation (linear in $\left(a/r_0\right)^2$) is obtained without the $\beta =
  6.0$ data. The ALPHA and Ward identity abscissae are slightly
  displaced for clarity.} 
\label{fig:Fpi2LANL}
\end{center}
\end{figure}

\parbreak
The above results have been obtained with the normalization constants
and improvement coefficients determined in various ALPHA Collaboration
publications. In order to monitor their influence on our data, we have
repeated the analysis using the LANL Collaboration results for the same 
quantities (for details, numerical values and references see
Appendix A of~\rep{Dimopoulos:2006dm}). The comparison of these
results, extrapolated to the physical kaon point, is displayed
in Fig.~\ref{fig:Fpi2LANL}. From it we draw the following conclusions:
\begin{itemize}
\item
The ALPHA and LANL results for each of the decay constants are
compatible, the only exceptions being $F_{sc}$ (incompatibility) and
$F_{sd}$ (near incompatibility) at $\beta = 6.0$. More detailed tests
indicate that the main source of incompatibility lies in
$\icA(\beta = 6.0)$.
\item
The ALPHA and LANL results for $F_{ud}$ are fully compatible to
those for $\tilde F_{ud}$. All three are independent from the axial current
(cf.~\req{eq:dc-ud} and~\req{eq:dc-ud2}) and thus from $\icA$. Moreover, $\tilde F_{ud}$
does not depend on any normalization constants and/or improvement
coefficients (cf.~\req{eq:dc-ud2}).
\item
Unlike the ALPHA results, at $\beta = 6.0$ the three LANL estimates
$F_{sc}$, $F_{sd}$ and $F_{ud}$ are fully compatible~\footnote{
The $\beta = 6.0$ LANL values at the physical kaon mass are
$r_0 F_{sc}=0.4205(63)$, $r_0 F_{sd}=0.4275(74)$ and $r_0 F_{ud}=0.4292(67)$.}
and show less scaling violations over the range of simulated couplings.
This implies that the $O(a^2)$ discrepancy shown in
Fig.~\ref{fig:isodecay} is significantly reduced if $\icA^{\rm LANL}$
is used instead of $\icA^{\rm ALPHA}$. Thus the tmQCD flavour breaking
effects under scrutiny, when monitored by decay constant ratios,
appear to be obscured by the uncertainty in the $\icA$ determination.
The mass ratios of Fig.~\ref{fig:isomass} are a more reliable
monitor of tmQCD flavour breaking.
\end{itemize}

\begin{figure}[ht]
\begin{center}
\epsfig{figure=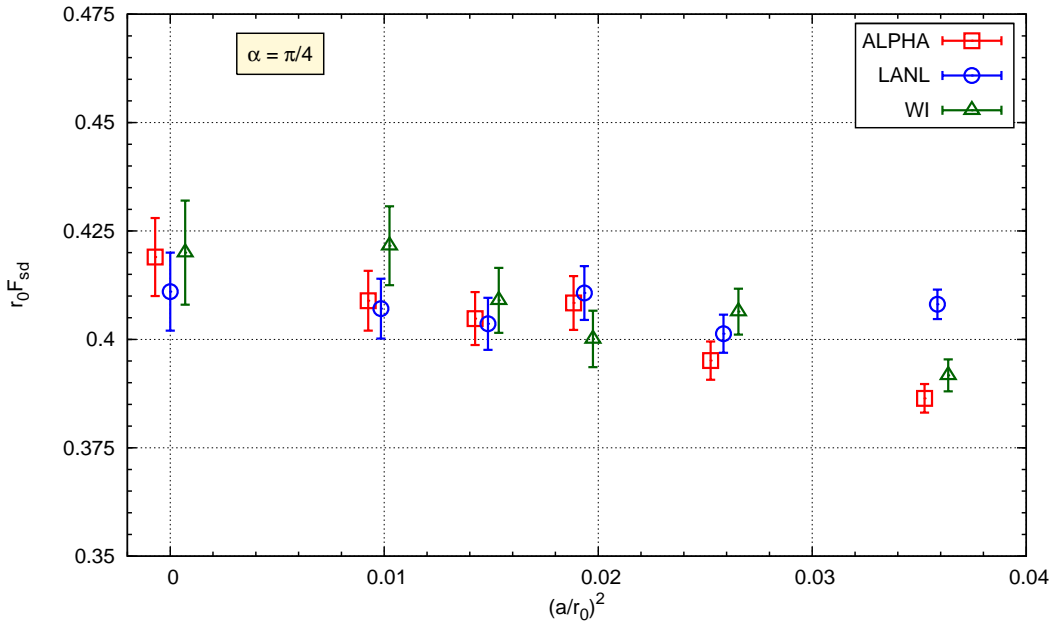, angle=-90, width=14.0 true cm}
\hskip 2.0cm
\caption{Pseudoscalar meson decay constants, obtained with ALPHA and
  LANL normalization and improvement coefficients at the physical kaon
  mass. The twist angle is $\pi/4$. The results for $r_0 \tilde F_{ud}$ 
  (based on a Ward identity) are also shown. The continuum
  extrapolation (linear in $\left(a/r_0\right)^2$) is obtained without the $\beta =
  6.0$ data. The ALPHA and Ward identity ordinates are slightly
  displaced for clarity.} 
\label{fig:Fpi4LANL}
\end{center}
\end{figure}

\parbreak
The continuum limit estimates for the various decay constants of the
$\pi/2$ case are obtained by linear extrapolation in $(a/r_0)^2$. 
Strictly speaking, Symanzik-improved quantities such as the decay
constants, contain some improvement coefficients which are only
known in perturbation theory (cf. $\tilde b_{\rm\scriptscriptstyle A}$ and 
$\tilde b_{\rm\scriptscriptstyle V}$).
This means that there are also $O(a g_0^4)$ discretization errors.
We have explicitly checked that the influence of the corresponding
counterterms is negligible in practice and therefore the dominant
discretization error is indeed $O(a^2)$. 
\parbreak
From Table~\ref{tab:decpi2} we see that our continuum limit results are
compatible across all flavour combinations considered. Moreover we note that
they do not change substantially if the point of the coarsest lattice 
($\beta = 6.0$) is removed.~\footnote{This is not in accordance with
  the findings of ref.~\cite{mbar:pap3} for $F_{sc}$, the continuum
  limit of which was obtained without the $\beta = 6.0$ result. This
  difference is explained by the bigger statistical sample ($O(1000)$
  configurations) and the simulations down to a finer lattice spacing 
  ($\beta =6.45$) of that work.}
From Fig.~\ref{fig:Fpi2LANL} we see that the continuum limit
extrapolations with ALPHA and LANL data are also compatible. We have
also confirmed that this conclusion remains valid if the $\beta = 6.0$
data are included in these extrapolations.
\parbreak
We now pass to the computation of the kaon masses and decay constants
in the $\pi/4$ case. The corresponding lattice correlation
function to be used in \req{eq:effmass} is:
\begin{itemize}
\item $f_{(\scrA_{\rm R})_{sd}}(x_0) \rightarrow
\dfrac{1}{\sqrt 2}[f_{(\scrA_{\rm R})_{sd}} (x_0) -i f_{(\scrV_{\rm R})_{sd}} (x_0)]$\ ,
\end{itemize}
where now flavours $s,d$ are both twisted. From it, the estimate
$M^{\rm eff}_{sd}$ is obtained. A second estimate $\tilde M^{\rm eff}_{sd}$
can also be obtained from the pseudoscalar density correlation
function $f_{(\scrP_{\rm R})_{sd}}$, with
\begin{equation}
(P_{\rm R})_{sd} = \ZP [ 1 + \ibP a m_{{\rm q},{\rm tw}} ] P_{sd}\ .
\end{equation}
The decay constant is obtained from the correlation function
\begin{equation}
F_{sd}  \approx
{\sqrt 2} (M^{\rm eff}_{sd} L^3)^{-1/2} \exp[
(x_0 - T/2)M^{\rm eff}_{sd}] \dfrac{f_{(\scrV_{\rm R})_sd}(x_0) - i f_{(\scrV_{\rm R})_sd}(x_0)}{\sqrt{f_{1,sd}}}\ .
\label{eq:dc-ud4}
\end{equation}
An alternative derivation is based on the continuum PCAC relation
\begin{equation}
\partial_\mu (A_{\rm R})_{\mu,sd} = 2 M_{\rm R, tw} (P_{\rm R})_{sd} = 2 {\sqrt 2} \muren{,{\rm tw}} (P_{\rm R})_{sd}\ .
\end{equation}
The last equation is derived by taking into consideration 
that~\req{quark_mass_M} reduces to $\ZP^{-1}  {\sqrt 2} \mu_{0,{\rm tw}}$ for degenerate
quark masses $\mren{,{\rm tw}} = \muren{,{\rm tw}}$. We thus obtain the decay
constant estimate
\begin{eqnarray}
\tilde F_{sd}  \approx - 4 {\sqrt 2}
\dfrac{\muren{,{\rm tw}}}{M^{\rm eff}_{sd}}
(M^{\rm eff}_{sd} L^3)^{-1/2} \exp[
(x_0 - T/2)M^{\rm eff}_{sd}] \dfrac{f_{(\scrP_{\rm R})sd}(x_0)}{\sqrt{f_{1,sd}}}\ .
\label{eq:dc-ud4pcac}
\end{eqnarray}
\parbreak
The results for the decay constants, obtained with 
ALPHA estimates for the normalization constants
and improvement coefficients, are collected in
Table~\ref{tab:decpi4}. In most cases there is again 
full compatibility between the masses
$r_0 M_{sd}^{\rm eff}$ and $r_0 \tilde M_{sd}^{\rm eff}$,
as well as the decay constants $r_0 F_{ud}$ and $r_0 \tilde F_{ud}$,
at all $\beta$ values. The comparison  with the LANL results,
made at the physical kaon mass, is displayed in
Fig.~\ref{fig:Fpi4LANL}. From it we draw the following conclusions:
\begin{itemize}
\item
The ALPHA and LANL results are compatible, except at $\beta = 6.0$. 
\item
Beyond $\beta = 6.1$, ALPHA and LANL results for $r_0 F_{sd}$ are compatible to
those for $r_0 \tilde F_{sd}$, the latter being independent of any
normalization constants and/or improvement coefficients (cf.~\req{eq:dc-ud4pcac}).
As the ALPHA $r_0 F_{sd}$ estimate scales like $r_0 \tilde F_{sd}$ for the whole $\beta$
range, the two have almost identical continuum limits.
\item
Compared to the ALPHA and Ward identity results, the LANL ones display a better scaling
behaviour over the whole range of simulated couplings.
\end{itemize}
\parbreak
The continuum limit extrapolations for the decay constants,
linear in $(a/r_0)^2$, are also
displayed in Table~\ref{tab:decpi4}. We see that they 
are all in good agreement and do not
change substantially if the point of the coarsest lattice 
($\beta = 6.0$) is removed. The same extrapolations (for
all $\beta$ values) performed with the LANL data yield
$r_0 F_{sd} = 0.404(6)$, which is incompatible to the ALPHA and Ward identity
estimates, due to its small error.
If however the $\beta = 6.0$ point is removed,
we obtain the compatible result $r_0 F_{sd} = 0.411(9)$,
which is the situation displayed in Fig.~\ref{fig:Fpi4LANL}.
\parbreak
The overall conclusion is that the continuum results are remarkably
stable for the different flavour combinations, and different
tmQCD regularizations (i.e. $\pi/2$ and $\pi/4$ cases). The best
result for $F_K$ in the continuum limit is obtained with a constrained
fit of $\tilde F_{ud}$ (for the $\pi/2$ case) and $\tilde F_{sd}$
(for the $\pi/4$ case). This choice is dictated by the absence of
(re)normalization and improvement coefficients in these quantities,
which amounts to the elimination of one source of systematic errors.
Our final $F_K$ estimate is
\begin{equation}
r_0 F_K = 0.421  \pm 0.007
\end{equation}
which agrees nicely with the previous ALPHA result
$r_0 F_K = 0.415  \pm 0.009$ of ref.~\cite{mbar:pap3},
and the $\chi$LF-Collaboration one $r_0 F_K = 0.410  \pm 0.011$
of ref.~\cite{Jansen:2005kk}.

\section{Four-fermion operators}
\label{sec:4ferm}

We now pass to the discussion of our tmQCD results concerning the
matrix elements of the parity-odd four-fermion operators
\begin{eqnarray}
{\cal Q}^\pm_1 = [ \bar \psi_1 \gamma_\mu \psi_2 ]
[ \bar \psi_3 \gamma_\mu \gamma_5 \psi_4 ] 
+ [ \bar \psi_1 \gamma_\mu \gamma_5 \psi_2 ]
[ \bar \psi_3 \gamma_\mu \psi_4 ] \pm [ 2 \leftrightarrow 4]
\end{eqnarray}
between pseudoscalar states. After attributing appropriate physical
quark flavours to the four-quark fields $\psi_k (k= 1,\cdots,4$),
we compute the ratios 
\begin{equation}
{\cal R}_\pm = \dfrac{\langle \pi^+ \vert {\cal Q}^\pm_1 \vert K^+ \rangle}
{\langle \pi^+ \vert A_{0,ud} \vert 0 \rangle 
\langle 0 \vert A_{0,su} \vert K^+ \rangle}\ .
\end{equation}
These ratios are the core quantities for the calculation of the kaon-to-pion
weak matrix elements related to the $\Delta I = 1/2$ rule. Note that
our computations do not directly provide the physical $K \rightarrow
\pi$ matrix elements of interest, as our pseudoscalar mesons (pions
and kaons) are degenerate and at best as light as the physical kaon.
Nevertheless our results have been used in
ref.~\cite{Dimopoulos:2006ma} in order to obtain the renormalization constants
of ${\cal Q}_1^\pm$ with Neuberger fermions. This is achieved through a
matching procedure involving the corresponding RGI
operators, computed with Wilson fermions. For details on the method
and the notation, see ref.~\cite{Dimopoulos:2006ma}. A different
relabelling of the operator quarks  $\psi_k (k= 1,\cdots,4$), allows
the identification of the matrix element of ${\cal Q}_1^+$ between
pseudoscalar states with
$\langle \bar K^0 \vert Q^{\Delta S =2} \vert K^0 \rangle$. Moreover,
in the quenched approximation we have the identification
$B_K = (3/4) {\cal R_+}$; see
refs.~\cite{Doninetal:4fRIMOM,ssf:vapav,Dimopoulos:2006dm} for details.
\vskip 0.2cm
Our aim here is twofold: First, we wish to report our $B_K$ results
at $\beta = 6.1$, computed with the new value of $\kappa_{\rm cr}$.
Second, we also list the values of ${\cal R_-}$ at all couplings;
the analysis of ref.~\cite{Dimopoulos:2006ma} was based on these results.
As the continuum extrapolation of ${\cal R_\pm}$
has also been performed in ref.~\cite{Dimopoulos:2006ma}, it needs
not to be repeated here. Nevertheless, we will discuss in some detail
the continuum extrapolation of $B_K$ (which is that of ${\cal R_+}$,
since $B_K = (3/4) {\cal R_+}$), as this amounts to
an update of our older $B_K$ result of ref.~\cite{Dimopoulos:2006dm}. 
This new $B_K$ analysis has also been presented in ref.~\cite{Pena:2006tw}.
\begin{figure}[ht]
\hskip 0.2cm\epsfig{figure=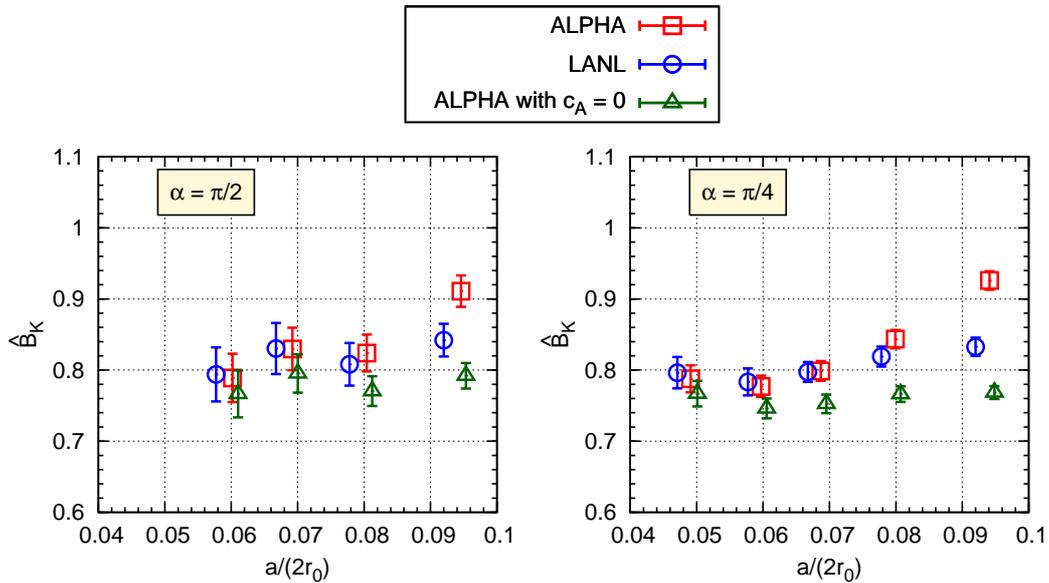, angle=-90, width=13.6 true cm}
\begin{center}
\caption{Uncertainties on $\hat B_K$ related to the $\Oa$ improvement of bilinears.
The left panel displays $\pi/2$ data, while the right panel shows $\pi/4$ data.}
\label{fig:bkvsa}
\end{center}
\end{figure}
\parbreak
All computational details are identical to those of ref.~\cite{Dimopoulos:2006dm}.
Following that work,
we construct three versions of the ratio ${\cal R_\pm}$, differing
in the structure of the $O(a)$ bilinear counterterm operators 
in its denominator. 
The results for ${\cal R_+}$ at $\beta = 6.1$, are collected in
Table~\ref{tab:resR+}; the corresponding RGI bag parameter $\hat B_K$  is
shown, for all lattice spacings, in Table~\ref{tab:BK} and
Fig.~\ref{fig:bkvsa}.
The determination of $\hat B_K$ in the continuum limit 
involves a linear extrapolation in $a$, with the two different 
regularisations ($\pi/2$ and $\pi/4$) combined in a fit constrained
to a common value at zero lattice spacing. It turned out that one of the most
relevant sources of cutoff effects is related to the arbitrariness
of the denominator $O(a)$ counterterms mentioned above. For instance,
using either the values for $\ZA,\ZV,\icA$ determined by the ALPHA Collaboration
or those obtained by the LANL group \cite{lat01:gupta} results 
in sizeable effects on $B_K$ at $\beta = 6.0$. At $\beta = 6.1$ we
also discern discretization effects in the $\pi/4$ case (see Figure~\ref{fig:bkvsa}). 
This signals the presence of large $O(a^2)$ ambiguities in $\hat B_K$ far from the
continuum limit. Combined linear+quadratic extrapolation of the data
proved to be unreliable, since the curvature of the quadratic term
dominates the result also close to the continuum limit. Linear fits,
excluding the $\beta = 6.0$ data, give
\begin{alignat}{3}
\hat B_K & = 0.668(45) \,, \qquad & & {\rm with} \,\, \icA^{\rm ALPHA}\ ; \\
\hat B_K & = 0.737(51) \,, \qquad & & {\rm with} \,\, \icA^{\rm LANL}\ ; \\
\hat B_K & = 0.744(42) \,, \qquad & & {\rm with} \,\, \icA = 0\ ;
\end{alignat}
whereas once also $\beta = 6.1$ is excluded we obtain
\begin{alignat}{3}
\hat B_K & = 0.735(71) \,, \qquad  & & {\rm with} \,\, \icA^{\rm ALPHA}\ ; \\
\hat B_K & = 0.771(80) \,, \qquad  & & {\rm with} \,\, \icA^{\rm LANL}\ ; \\
\hat B_K & = 0.780(67) \,, \qquad  & & {\rm with} \,\, \icA = 0\ .
\end{alignat}
In the $\pi/2$ case, the error decrease in the extrapolations upon
including the $\beta=6.1$ point (for which LANL and ALPHA
data are fully compatible), is marginal. For instance, combined  linear
extrapolations with $\icA^{\rm ALPHA}$, excluding only
the $\beta=6.0$ data for $\pi/2$ and those at
$\beta=6.0, 6.1$ for $\pi/4$, yield $\hat B_K = 0.739(66)$.
\parbreak
The above results indicate that the extrapolation of the LANL data 
is the most stable. In spite of this, the ALPHA data is considered
to be the best estimate, on grounds related to the systematic uncertainties
in the derivation of $\ZA$, $\ZV$ and $\icA$, as
explained at length in ref.~\cite{Dimopoulos:2006dm}. Since 
the difference between ALPHA and LANL results is only significant
at $\beta=6.0$ and $\beta=6.1$, we have conservatively discarded
these data points in the continuum extrapolation, illustrated in
the left panel of Figure~\ref{fig:bk}. The final results are:
\begin{align}
\label{eq:BKRGI}
\hat B_K & = 0.735(71) \,,\\
B_K^{\MSbar}(2~\GeV) & = 0.534(52) \,.
\end{align}
When comparing with the result ($\hat B_K = 0.789(46)$) quoted in
\cite{Dimopoulos:2006dm}, we must take into account that the revision of
the $\beta=6.1$ data has important consequences for the continuum
limit extrapolation. In the analysis of
ref.~\cite{Dimopoulos:2006dm} (cf. right panel of Figure~\ref{fig:bk}),
good scaling behaviour appeared to set-in rather abruptly at $\beta~=~6.1$,
with continuum limit extrapolation becoming stable only once the
$\beta=6.0$ points were discarded. On the contrary, the new $\beta=6.1$
data interpolate in a smoother way those at $\beta=6.0$ and $\beta=6.2$
(cf. left panel of Figure~\ref{fig:bk}). This however, implies a worsening
of the scaling behaviour, with continuum limit extrapolation becoming
stable only upon discarding our $\beta=6.0$ and $\beta=6.1$ results, as
detailed above.
\vskip 0.0cm
\begin{figure}[!ht]
\epsfig{figure=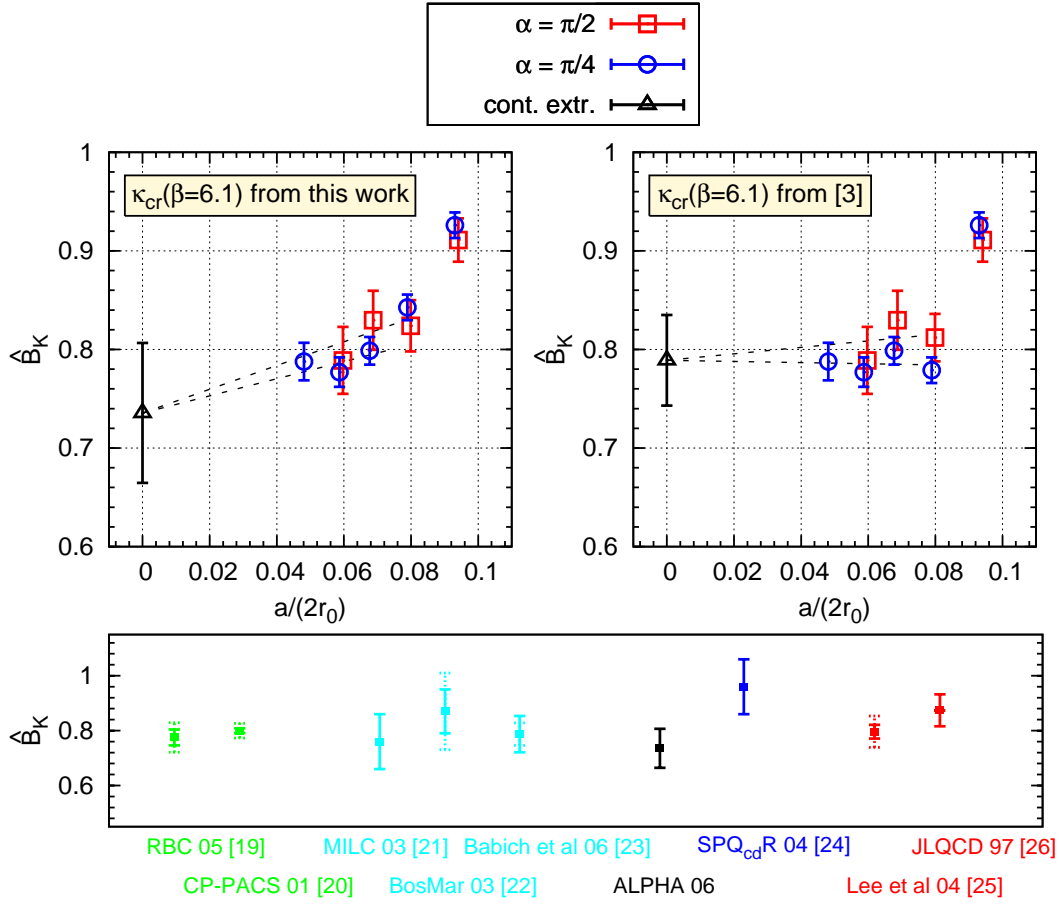, angle=-90, width=13.6 true cm}
\vskip -0.5cm
\begin{center}
\caption{Left: Continuum limit extrapolation of $\hat B_K$. The
  $\pi/2$ ordinates have been slightly displaced for clarity. Right:
  Same as before, but with $\hat B_K(\beta=6.1)$ from \cite{Dimopoulos:2006dm}.
  Bottom: Comparison with other quenched results; different fermion discretizations 
  are as follows (from left to right): \cite{Aoki:2005ga,AliKhan:2001wr} 
  domain wall; \cite{DeGrand:2003in,Garron:2003cb,Babich:2006bh} overlap; 
  \cite{Becirevic:2004aj} Wilson; \cite{Lee:2004qv,Aoki:1997nr} staggered. 
  The ALPHA point is that of the present work (Wilson tmQCD). }
\label{fig:bk}
\end{center}
\end{figure}
\parbreak
The value of $\hat B_K$ is shown in
the bottom panel of Figure~\ref{fig:bk} alongside
other representative results in quenched QCD found in the literature
\cite{Aoki:2005ga,AliKhan:2001wr,DeGrand:2003in,Garron:2003cb,Babich:2006bh,
Becirevic:2004aj,Lee:2004qv,Aoki:1997nr}. 
Our result is the only quenched result which has simultaneously 
eliminated the systematic uncertainties related to
renormalisation (both at a reference scale and from the point of view
of RG running), ultraviolet cutoff dependences, and finite volume effects
(within the available accuracy). On the other hand, the control of the
mass dependence of $\hat B_K$ with Wilson fermions is still not as
accurate as with e.g. Neuberger or domain wall fermions. 
\parbreak
The ${\cal R}_-$ results for all $\beta$ values are displayed in
Tables~\ref{tab:respi2_minus} and~\ref{tab:respi4_minus}. The extrapolation
of this data to the continuum limit has been presented in ref.~\cite{Dimopoulos:2006ma}.

\section{Conclusions}
\label{sec:conclusions}

In this work we have completed our study of basic kaon weak matrix elements
in quenched Wilson tmQCD.
Our final value for $F_K$ is the best controlled quenched result obtained 
with Wilson fermions from the point of view of systematic uncertainties. 
We have also provided a final value for $B_K$, with an error that, in 
our view, reflects faithfully the best accuracy that can be expected 
for this quantity in the absence of full $\Oa$ improvement.
Finally, we have performed a thorough study of flavour breaking in
our version of tmQCD, confirming that it does not introduce
any uncontrolled systematics in our results.
\parbreak
The dominant source of uncertainty left in the quenched approximation
(certainly so for $B_K$) is related to the lack of full $\Oa$ improvement, which
amplifies the error of the continuum limit extrapolation. Thus, if Wilson
fermions are to be used in the future in the determination of weak matrix
elements, the use of tmQCD variants that embody automatic $\Oa$ improvement
\cite{Frezzotti:2004wz} may prove essential. Two important aspects of the tmQCD
approach are crucial in the context of weak matrix elements: the tuning of
mass parameters, in particular of the twist angle, has to be controlled to high
precision; and flavour symmetry breaking effects should be reasonably small, 
as in the present study. In conclusion, the present work
demonstrates that, once the tuning of the twisted angle and flavour symmetry
breaking are under control, tmQCD may offer a convenient alternative to other 
discretizations. As we are entering the era of tmQCD simulations with 
dynamical quarks \cite{Boucaud:2007uk}, it will be important to explore 
these issues in future unquenched studies.

\section*{Acknowledgments}
We wish to thank H.~Wittig for help and discussions.
A.V. wishes to thank R.~Frezzotti and G.C.~Rossi for discussions.
F.P. acknowledges financial support from the
Alexander-von-Humboldt Stiftung.
C.P. acknowledges partial financial support by CICyT project FPA2006-05807.
We wish to thank CERN, DESY-Zeuthen and INFN-Rome2
for providing hospitality to several
members of our collaboration at various stages of this work.
We also thank the DESY-Zeuthen computing centre for its support.
This work was supported in part by the EU Contract 
No. MRTN-CT-2006-035482, "FLAVIAnet".
\begin{appendix}
\section{The effect of an offset in the $\kappa_{\rm cr}$ estimate}
\label{sec:appkcrit}

In this appendix we present the results for $\cot(\taa)$, the
pseudoscalar effective mass and the decay constant, computed
with quark mass parameters tuned with the value of $\kappa_{\rm cr}(\beta=6.1)$
quoted in ref.~\cite{mbar:charm1}. As shown in  Table~\ref{tab:kcrit},
this value, obtained by interpolation over a range of
$\kappa_{\rm cr}(\beta)$, is roughly only $0.1\%$ off the estimate
of the present work, computed directly at $\beta = 6.1$.
This apparently small offset is nevertheless 
a discrepancy of 15 standard deviations and has significant
consequences in the quantities of interest, which we now examine in
some detail.
\parbreak
We denote by $\kappa_{\rm cr}$ the value computed on the constant physics condition of
ref.~\cite{impr:pap3}; at $\beta = 6.1$ this computation has been performed
in the present work. The estimate at $\beta = 6.1$, obtained through
interpolation of data computed at several other gauge
couplings~\cite{mbar:charm1}, is parametrized as
$\kappa_{\rm cr}^\prime = \kappa_{\rm cr}[1+\delta_{\rm cr}]$. Let us keep
track of the effect of this offset in the tuning of the various quark masses, starting
with the $\pi/2$ case. The tuning of the hopping parameter $\kappa_{\rm w}$ 
of the untwisted doublet is done in our simulations by requiring that the
pseudoscalar meson made of the two untwisted flavours, has a mass $M^{\rm eff}_{sc}$ 
equal to a given value, fixed between $640~\MeV$ and $830~\MeV$. This
procedure is unaffected by $\kappa_{\rm cr}$, as is the computation of purely
untwisted quantities such as $r_0 F_{sc}$. What is affected by the offset 
$\delta_{\rm cr}$ is our estimate of the subtracted masses, which now become
\begin{equation}
a m_{{\rm q},f}^\prime = \dfrac{1}{2\kappa_f} - \dfrac{1}{2 \kappa_{\rm cr}^\prime}
= a m_{{\rm q},f} + (\delta a m_{\rm cr}) \, , \qquad (f={\rm tw},{\rm w})\ ,
\label{eq:dmq}
\end{equation}
with 
\begin{equation}
(\delta a m_{\rm cr})    \equiv \delta_{\rm cr} / (2 \kappa_{\rm cr}) + \cdots
\,\, ,
\end{equation}
where terms of $O(\delta_{\rm cr}^2)$ have been dropped. So now~\req{eq:msren}
becomes
\begin{eqnarray}
a \mren{,{\rm w}}^\prime = a \mren{,{\rm w}} + \Zm (\delta a m_{\rm cr}) + \cdots
  \,\, .
\end{eqnarray}
where $O(a)$ counterterms are dropped, when multiplied by $\delta a m_{\rm cr}$.
The requirement of quark mass degeneracy now reads $\mren{,{\rm w}}^\prime =
\muren{,{\rm tw}}^\prime$;
i.e. the offset filters through to the twisted mass, which now becomes
\begin{eqnarray}
a \muren{,{\rm tw}}^\prime = a \muren{,{\rm tw}} + \Zm (\delta a m_{\rm cr}) + \cdots\ .
\label{eq:muprime}
\end{eqnarray} 
This clearly induces an offset $\ZP^{-1}  \Zm (\delta a m_{\rm cr})$
in the tuning of the bare twisted
mass parameter, which we therefore denote as $a \mu_{0,{\rm tw}}^\prime$.
Finally, the hopping parameter of the twisted doublet is tuned to a value
$\bar \kappa_{\rm tw}$, corresponding to 
\begin{eqnarray}
a \bar m_{{\rm q},{\rm tw}}^\prime = \dfrac{1}{2 \bar \kappa_{\rm tw}} 
- \dfrac{1}{2 \kappa_{\rm cr}^\prime}
= a \bar m_{{\rm q},{\rm tw}} + (\delta a m_{\rm cr}) 
\,\, ,
\label{eq:mbarql}
\end{eqnarray} 
by requiring the vanishing of the renormalized light quark mass,
which is now written as
\begin{eqnarray}
a \bar m_{{\rm R},{\rm tw}}^\prime = \Zm [ a \bar m_{{\rm q},{\rm tw}}^\prime 
(1+\ibm a \bar m_{{\rm q},{\rm tw}}^\prime)
+ \tilde b_{\rm m} (a \mu_{0,\rm tw}^\prime)^2 ] \,\, .
\label{eq:mrenprime}
\end{eqnarray} 
It is important to note that the above quantity is not the
true standard quark mass in the twisted sector.
Considering that $\kappa_{\rm cr}$ (and not $\kappa_{\rm cr}^\prime$) is
by definition the reliably estimated critical point, the true
renormalized quark mass, for the  hopping parameter $\bar \kappa_{\rm tw}$,
is expressed in terms of $\bar m_{{\rm q},{\rm tw}}^\prime$ (cf. \req{eq:mbarql})
as $a\bar m_{{\rm q},{\rm tw}} = a \bar m_{{\rm q},{\rm }}^\prime - (\delta a m_{\rm cr})$.
It turns out to be non-zero:
\begin{eqnarray}
a \bar m_{{\rm R},{\rm tw}} = \Zm [ a \bar m_{{\rm q},{\rm tw}} ( 1 + \ibm a
  \bar m_{{\rm q},{\rm tw}}) + \tilde b_{\rm m} (a
  \mu^\prime_{0,\rm tw})^2 ] = - \Zm (\delta a m_{\rm cr}) \ .
\label{eq:mqhatprime}
\end{eqnarray}
The second expression has been derived by implementing the vanishing of
\req{eq:mrenprime}.
\parbreak
The bottom line is that we now have a theory characterized by
the bare parameters $\kappa_{\rm w}, \bar \kappa_{\rm tw}, \mu_{\rm tw}^\prime$
(or, equivalently, $m_{{\rm q},\rm w}, \bar m_{{\rm q},\rm tw}, \mu_{0,\rm tw}^\prime$) which correspond
to the same heavy quark mass $\mren{,{\rm w}}$, a small but non-zero
light quark mass $\bar m_{{\rm R},{\rm tw}}$ and a twisted mass
$\muren{,{\rm tw}}^\prime$.
A twist angle defined through the ratio of mass parameters
$a \muren{,{\rm tw}}^\prime$ and $a \bar m_{{\rm R},{\rm tw}}^\prime$, is tautologically equal to
$\pi/2$. Instead, the true twist angle of the theory is given by
\begin{eqnarray}
\tan (\bar \alpha^\prime) =
\dfrac{a\muren{,{\rm tw}}^\prime}{a \bar m_{{\rm R},{\rm tw}}} =
\dfrac{a\mren{,{\rm w}}^\prime}{a \bar m_{{\rm R,{\rm tw}}}}
= - \dfrac{a m_{{\rm q},{\rm w}}}{\delta a m_{\rm cr}} + \cdots\ ,
\end{eqnarray}
which may differ significantly from the target value $\pi/2$.
\parbreak
In Table~\ref{tab:cota61}, we show the values of $\cot(\taa)$.
It is clear that they are completely incompatible to the target
value of $\pi/2$ (and $\pi/4$; see below), on which the mass parameter 
tuning is based. This is in contrast to the small deviation from the 
target values of the $\cot(\taa)$ estimates given in Table~\ref{tab:cota}, 
which simply reflects the presence of $O(a^2)$ effects.
\parbreak
In Table~\ref{tab:mf61pi2}, the results for the various pseudoscalar effective 
masses and decay constants are presented for the $\pi/2$ case. The quantities 
$r_0 M^{\rm eff}_{sc}$ and $r_0 F_{sc}$ are not reported here, as they
are identical to those of Tables~\ref{tab:meffpi/2}
and~\ref{tab:decpi2}. This is because they
consist of untwisted flavours, the mass tuning of which is
independent of $\kappa_{\rm cr}$. 
\parbreak
There is a rough check which enables us to ``predict'' the discrepancy
in our results, induced by an offset $\delta \kappa_{\rm cr}$ in the
critical point.
The standard PCAC dependence of the squared pseudoscalar mass on
the average valence quark masses implies that:
\begin{align}
\left(a M^{\rm eff}_{sc} \right)^2 & \propto 2 \,\, a \mren{,{\rm w}}\ , \\
\left(a M^{\rm eff}_{ud} \right)^2 & \propto 2 [ a \mu_{\rm R,{\rm tw}} + 
\Zm (\delta a m_{\rm cr}) ]\ , \\
\left( a M^{\rm eff}_{sd} \right)^2 & \propto a \mren{,{\rm w}} + a
\muren{,{\rm tw}} + \Zm
(\delta a m_{\rm cr})\ .
\end{align}
The above expressions have been obtained by keeping track of the offset
in the various quark masses (cf. Eqs.~(\ref{quark_mass_M}),
(\ref{eq:muprime}) and (\ref{eq:mqhatprime})) in the tuning procedure,
through straightforward lowest order Taylor expansions in
$\delta a m_{\rm cr}$. Since, in the absence of offset $\delta a m_{\rm cr}$,
the quark masses are tuned to satisfy $a \mren{,{\rm w}} = a
\muren{,{\rm tw}}$, 
we easily derive
\begin{align}
\left( \dfrac{a M^{\rm eff}_{sd}}{a M^{\rm eff}_{sc}} \right)^2
& = 1 + \dfrac{\Zm (\delta a m_{\rm cr})}{2 a \mren{,{\rm w}}}\ ,
\\
\left(\dfrac{a M^{\rm eff}_{ud}}{a M^{\rm eff}_{sc}} \right)^2
& = 1 + \dfrac{\Zm (\delta a m_{\rm cr})}{a \mren{,{\rm w}}} \,\, .
\label{eq:massratFB}
\end{align}
This means that the offset of the first ratio, due to
$\delta \kappa_{\rm cr}$, is ``predicted'' to be
half of that of the second. 
This is roughly confirmed by the data in the case of
$\beta = 6.1$ with the offset $\kappa_{\rm cr}$. The relevant results
are gathered in Table~\ref{tab:respi2ratios}. Clearly, as the
whole procedure does not take into account higher order
discretization effects, our expectations are confirmed
at a qualitative level.
\parbreak
We now turn to the $\pi/4$ case, in which the mass tuning proceeds
in a different way (cf. ref.~\cite{Dimopoulos:2006dm}).
For a fixed bare twisted mass $a \mu_{0,{\rm tw}}$,
the mass degeneracy condition $\mren{,{\rm tw}} = \muren{,{\rm tw}}$ 
fixes the subtracted mass $m_{{\rm q},{\rm tw}}$,
in terms of Eqs.~(\ref{eq:mlrenu}) and (\ref{eq:mlren}), to the value
\begin{equation}
a m_{{\rm q},{\rm tw}} = \dfrac{1}{Z \ZA} a \mu_{0,{\rm tw}} \left\{ 1 +
\left[ \dfrac{1}{Z \ZA} (b_\mu - \ibm ) - Z \ZA \tilde b_{\rm m} \right]
a \mu_{0,{\rm tw}} \right\}\ .
\end{equation} 
Now this value of $a m_{{\rm q},{\rm tw}}$ induces fixing the hopping
parameter to say, $\kappa$ or $\bar \kappa$, depending on whether
we are working with $\kappa_{\rm cr}$ or $\kappa_{\rm cr}^\prime$:
\begin{equation}
a m_{{\rm q},{\rm tw}} = \dfrac{1}{2 \kappa} -  \dfrac{1}{2 \kappa_{\rm cr}}
= \dfrac{1}{2 \bar \kappa} -  \dfrac{1}{2 \kappa_{\rm cr}^\prime}\ .
\end{equation} 
When we perform simulations at hopping parameter $\bar\kappa$ (based on
the tuning with $\kappa_{\rm cr}'$), we are not really at  subtracted quark mass
$a m_{{\rm q},{\rm tw}}$, but rather at $a \bar m_{{\rm q},{\rm tw}} =
1/(2\bar\kappa)-1/(2\kappa_{\rm cr})$. This implies
the following offset in the subtracted quark mass:
\begin{equation}
a \bar m_{{\rm q},{\rm tw}} = a m_{{\rm q},{\rm tw}} + \delta a m_{\rm cr}\ .
\end{equation} 
The true untwisted quark masses of our simulation are then
\begin{eqnarray}
a \bar m_{{\rm R},{\rm tw}} &=& a \mren{,{\rm tw}} - \Zm \delta a
m_{\rm cr} + \cdots\ , \\
a \bar \mu_{{\rm R},{\rm tw}} &=& a \muren{,{\rm tw}} + \cdots\ ,
\end{eqnarray} 
with higher orders omitted. Combining these two expressions, we 
finally arrive at the estimate for the twist angle
\begin{equation}
\cot(\taa) = 1 - \dfrac{\Zm \delta a m_{\rm cr}}{a \muren{,{\rm tw}}}\ .
\label{eq:taap4off}
\end{equation} 
This is again the source of a significant deviation from the target
twist angle $\pi/4$; the Ward identity results of Table~\ref{tab:cota61}
discussed above corroborate this conclusion.
\parbreak
In Table~\ref{tab:mf61pi4}
we list the results for the pseudoscalar masses and decay constants.
They are significantly different to the ones obtained with
the new $\kappa_{\rm cr}(\beta=6.1)$ (cf. Table~\ref{tab:decpi4}).
We also notice that the excellent agreement between $r_0 M_{sd}^{\rm eff}$
and $r_0 \tilde M_{sd}^{\rm eff}$ in Table~\ref{tab:decpi4} is lost in
Table~\ref{tab:mf61pi4}. Recalling that the two quantities in question,
essentially being the two sides of a Ward identity, are equal up to discretization
effects, we interpret this discrepancy as a signal of $O(a^2)$
flavour symmetry violations. The comparison of $r_0 F_{sd}$ to
$r_0 \tilde F_{sd}$ confirms these conclusions.
\parbreak
The previous analysis of the mass offsets in the $\pi/4$ case
suggests two rough checks of the observed discrepancies. First,
we note that the twist angle ``prediction'' of \req{eq:taap4off}
gives, for the three twisted bare masses used,
\begin{eqnarray}
\cot(\taa) &\sim& 1.50\ , \\
\cot(\taa) &\sim& 1.54\ , \\
\cot(\taa) &\sim& 1.64\ ,
\end{eqnarray} 
which is in good qualitative agreement with the Ward identity estimates
listed in Table~\ref{tab:mf61pi4}. Second, we compare the pseudoscalar
effective masses $M^{\rm eff}_{sd}$, computed with our hopping
parameter $\kappa$ (tuned with $\kappa_{\rm cr}$) and listed in 
Table~\ref{tab:decpi4}, to the ones computed with the hopping
parameter $\bar \kappa$ (tuned with $\kappa_{\rm cr}^\prime$) of 
Table~\ref{tab:mf61pi4}. Henceforth, the latter quantities are
denoted as $\bar M^{\rm eff}_{sd}$. PCAC suggests that
\begin{eqnarray}
aM^{\rm eff}_{sd} &\propto& 2 aM_{{\rm R}, {\rm tw}}\ , \\
a\bar M^{\rm eff}_{sd} &\propto& 2 a\bar M_{{\rm R}, {\rm tw}} = 
2 \sqrt{ a\bar m_{{\rm R},{\rm tw}}^2 + a\bar \mu_{{\rm R},{\rm tw}}^2 }
= 2 aM_{{\rm R}, {\rm tw}} \Big[ 1 - 
\dfrac{\Zm \delta a m_{\rm cr}}{2 a\mu_{{\rm R},{\rm tw}}} \Big].
\label{eq:massratFB4}
\end{eqnarray}
A glance at Table~\ref{tab:respi4ratios} shows that the agreement
between the measured quantity $\left(a\bar M^{\rm eff}_{sd}/aM^{\rm eff}_{sd}
-1\right)$ and the ``predicted'' value $\Zm \delta a m_{\rm cr}/2 a\mu_{{\rm R},{\rm tw}}$
exceeds expectations.

\newpage

\section{Tables}
\label{sec:apptables}

\begin{table}[!ht]
\begin{center}
\begin{tabular}{ccccccc}
\Hline \\[-10pt]
$\beta$ & $(\frac{L}{a})^3 \times \frac{T}{a}$ & $\frac{a}{2 r_0}$ & $\frac{L}{2 r_0}$  & $\kappa_{\rm w}$ & 
$(\kappa_{\rm tw},a\mu_{0,{\rm tw}})$ & $N_{\rm conf}$\\ \\[-10pt]
\Hline \\[-10pt]
6.0 & $16^3 \times 48$ & 0.0931 & 1.49 & 0.1335 & (0.135169,0.03816) & 402 \\
& & & & 0.1338 & (0.135178,0.03152) & 398 \\
& & & & 0.1340 & (0.135183,0.02708) & 402 \\
& & & & 0.1342 & (0.135187,0.02261) & 400 \\ \\[-10pt]
\hline\\[-10pt]
6.1 & $24^3 \times 56$ & 0.0789 & 1.89 & 0.1343 & (0.1356465, 0.031711) & 100 \\
& & & & 0.1345 & (0.1356510, 0.027123) & 100 \\
& & & & 0.1347 & (0.1356560, 0.022523) & 122 \\ \\[-10pt]
\hline\\[-10pt]
6.2 & $24^3 \times 64$ & 0.0677 & 1.63 & 0.1346 & (0.1357800,0.0283240) & 200 
\\
& & & & 0.1347 & (0.1357825,0.0259850) & 201 \\
& & & & 0.1349 & (0.1357866,0.0212897) & 214 \\ \\[-10pt]
\hline\\[-10pt]
6.3 & $24^3 \times 72$ & 0.0587 & 1.41 & 0.1348 & (0.1358118,0.0246230) & 200 
\\
    &                  &       &      & 0.1349 & (0.1358139,0.0222430) & 205 
\\
    &                  &       &      & 0.1350 & (0.1358157,0.0198558) & 175 
\\
    &                  &       &      & 0.1351 & (0.1358174,0.0174640) & 201 
\\ \\[-10pt]
\Hline
\end{tabular}
\end{center}
\caption{The parameters of the run at twist angle $\taa = \pi/2$. The
  dataset at $\beta = 6.1$ is a new run, while all other data
  are those of ref.~\cite{Dimopoulos:2006dm}.
}
\label{tab:runspi2}
\end{table}

\begin{table}[!htp]
\begin{center}
\begin{tabular}{ccccccc}
\Hline \\[-10pt] 
$\beta$ & $(\frac{L}{a})^3 \times \frac{T}{a}$ & $\frac{a}{2 r_0}$ & $\frac{L}{2 r_0}$ & $\kappa_{\rm tw}$ & $a \mu_{0,{\rm tw}}$ & $N_{\rm conf}$\\ \\[-10pt]
\Hline \\[-10pt]
6.0 & $24^3 \times 48$ & 0.0931 & 2.24 & 0.134739 & 0.010412 & 200 \\
& & & & 0.134795 & 0.009142 & \\
& & & & 0.134828 & 0.008397 & \\ \\[-10pt]
\hline \\[-10pt]
6.1 & $24^3 \times 60$ & 0.0789 & 1.89 & 0.135320 & 0.00810 & 196 \\
& & & & 0.135358 & 0.00720 & \\
& & & & 0.135403 & 0.00615 & \\ \\[-10pt]
\hline \\[-10pt]
6.2 & $32^3 \times 72$ & 0.0677 & 2.17 & 0.135477 & 0.007595 & 73 \\
& & & & 0.135539 & 0.006125 & \\ \\[-10pt]
\hline \\[-10pt]
6.3 & $32^3 \times 72$ & 0.0587 & 1.88 & 0.135509 & 0.0076 & 76 \\
& & & & 0.135546 & 0.0067 & \\
& & & & 0.135584 & 0.0058 & \\ \\[-10pt]
\hline \\[-10pt]
6.45 & $32^3 \times 86$ & 0.0481 & 1.54 & 0.135105 & 0.01459 & 105 \\
& & & & 0.135218 & 0.01185 & \\
& & & & 0.135293 & 0.01002 & \\ \\[-10pt]
\Hline
\end{tabular}
\end{center}
\caption{The parameters of the run at twist angle $\taa = \pi/4$. The
  dataset at $\beta = 6.1$ is a new run, while all other data
  are those of ref.~\cite{Dimopoulos:2006dm}.
}
\label{tab:runspi4}
\end{table}

\begin{table}[!htp]
\begin{center}
\begin{tabular}{cll}
\Hline \\[-10pt] 
$\beta$ & \ \ \ \ $\kappa_{\rm cr}$ & ref. \\ \\[-10pt]
\Hline \\[-10pt]
6.0  & 0.135196(14)   & \cite{impr:pap3} \\
6.1  & 0.135496,      & \cite{mbar:charm1} \\
     & 0.135665(11)   & this work \\
6.2  & 0.135795(13)   & \cite{impr:pap3} \\
6.3  & 0.135823       & ZeRo Coll. \\
6.4  & 0.135720(9)    & \cite{impr:pap3} \\
6.45 & 0.135701       & \cite{mbar:charm1} \\ \\[-10pt]
\Hline
\end{tabular}
\end{center}
\caption{The critical hopping parameter $\kappa_{\rm cr}$ at several
  $\beta$-values, obtained from several sources. At $\beta = 6.1$ a
  second estimate has been computed in the present work. When
  available, errors have been quoted.}
\label{tab:kcrit}
\end{table}

\newpage

\begin{table}[!htp]
\begin{center}
\begin{tabular}{cccccc}
\Hline \\[-10pt] 
$\beta$ & $a \muren{,{\rm tw}}$ & $a \muren{,{\rm tw}}$ & $a \mren{,{\rm tw}}$ & $\cot(\taa)$ &
$\cot(\taa)$ \\ \\ [-10pt]
& \req{eq:mlren} & \req{eq:mrenpcvc} & \req{eq:mrenpcac} & \req{eq:tanapcac} & \req{eq:tanapcvc} \\ \\[-10pt]
\Hline \\[-10pt]
6.0 &  0.07292& 0.07458(2)& 0.00427(15)&    0.059 (2) & 0.054 (2) \\
     &  0.06016& 0.06129(2)& 0.00255(15)&    0.042 (3) & 0.039 (2) \\
     &  0.05176& 0.05245(2)& 0.00141(16)&    0.027 (3) & 0.025 (3) \\
     &  0.04321& 0.04360(2)& 0.00026(16)&    0.006 (4) & 0.004 (4) \\ \\[-10pt]
\hline \\[-10pt]
6.1 &  0.06097& 0.06222(2)& 0.00269(13)&   0.044 (2) & 0.041 (2) \\
     &  0.05216& 0.05305(2)& 0.00176(13)&   0.034 (2) & 0.031 (2) \\
     &  0.04332& 0.04395(2)& 0.00122(12)&   0.028 (3) & 0.026 (3) \\ \\[-10pt]
\hline \\[-10pt]
6.2 &  0.05470& 0.05545(1)& 0.00043(09)&  0.008 (2) & 0.006 (2) \\
     &  0.05018& 0.05080(1)& 0.00025(09)&  0.005 (2) & 0.003 (2) \\
     &  0.04112& 0.04152(1)& -0.00017(08)& -0.004 (2) &-0.006 (2) \\ \\[-10pt]
\hline \\[-10pt]
6.3 &  0.04769& 0.04833(1)& 0.00109(07)&  0.023 (2) & 0.021 (2) \\
     &  0.04308& 0.04364(1)& 0.00100(08)&  0.023 (2) & 0.022 (2) \\
     &  0.03846& 0.03890(1)& 0.00072(09)&  0.019 (2) & 0.017 (2) \\
     &  0.03382& 0.03418(1)& 0.00054(08)&  0.016 (2) & 0.015 (2) \\ \\[-10pt]
\hline\hline \\[-10pt]
6.0 & 0.01986 & 0.020091(3)& 0.01959(11)&   0.980(6) & 0.958(5)  \\
     & 0.01744 & 0.017646(3)& 0.01734(11)&   0.989(7) & 0.968(6)  \\
     & 0.01602 & 0.016210(3)& 0.01601(11)&   0.994(7) & 0.974(7) \\ \\[-10pt]
\hline \\[-10pt]
6.1 & 0.01555 & 0.015718(2)& 0.01482(10)&   0.948(3) & 0.933(6) \\
     & 0.01383 & 0.013974(2)& 0.01317(10)&   0.949(7) & 0.934(7) \\
     & 0.01181 & 0.011938(2)& 0.01120(11)&   0.945(9) & 0.931(9) \\ \\[-10pt]
\hline \\[-10pt]
6.2 &  0.01465& 0.014780(2)& 0.01540(8)&   1.047(5) & 1.034(5) \\
     &  0.01182& 0.011922(2)& 0.01261(8)&   1.064(7) & 1.051(7) \\ \\[-10pt]
\hline \\[-10pt]
6.3 &  0.01470& 0.014809(2) &0.01436(9) &  0.973(6) & 0.963(6) \\
     &  0.01296& 0.013056(2) &0.01266(9) &  0.973(7) & 0.964(7) \\
     &  0.01122& 0.011303(2) &0.01091(10) &  0.970(9) & 0.960(8) \\ \\[-10pt]
\hline \\[-10pt]
6.45 &  0.02823& 0.028417(2) &0.02733(5) &  0.962(2) & 0.951(2) \\
     &  0.02294& 0.023085(2) &0.02203(5) &  0.956(2) & 0.946(2)  \\
     &  0.01940& 0.019523(1) &0.01852(5) &  0.950(2) & 0.941(2) \\ \\[-10pt]
\Hline
\end{tabular}
\end{center}
\caption{Renormalized quark masses $a \mu_{{\rm R},{\rm tw}}$
and $a m_{{\rm R},{\rm tw}}$ and twist angles, computed from different procedures,
indicated by the corresponding equations. Upper table: data for
the $\pi/2$ case; lower table: data for the $\pi/4$ case.
}
\label{tab:cota}
\end{table}

\begin{table}[!htp]
\begin{center}
%\scriptsize
\begin{tabular}{cccccc}
\Hline \\[-10pt]
$\beta$ & $a \mu_{0,{\rm tw}}$ & $r_0M^{\rm eff}_{sc}$ & $r_0M^{\rm eff}_{sd}$ & 
$r_0M^{\rm eff}_{ud}$ & $r_0 \tilde M^{\rm eff}_{ud}$ \\ \\[-10pt]
\Hline \\[-10pt]
6.0 & 0.03816 & 2.089(6) & 2.092(6) & 2.054(5) & 2.054(5) \\
    & 0.03152 & 1.900(7) & 1.907(7) & 1.865(6) & 1.865(5) \\
    & 0.02708 & 1.771(7) & 1.780(6) & 1.738(5) & 1.733(5) \\
    & 0.02261 & 1.620(7) & 1.635(6) & 1.594(5) & 1.587(5) \\ \\[-10pt]
\hline \\[-10pt]    
6.1 & 0.0317110 & 2.024(6) & 2.039(6) & 2.027(5) & 2.023(6) \\
    & 0.0271230 & 1.858(7) & 1.874(7) & 1.861(7) & 1.860(6) \\
    & 0.0225230 & 1.693(7) & 1.715(6) & 1.703(6) & 1.700(6) \\ \\[-10pt]
\hline \\[-10pt]    
6.2 & 0.0283240 & 2.080(6) & 2.079(6) & 2.062(6) & 2.060(5) \\
    & 0.0259850 & 1.981(7) & 1.980(7) & 1.964(6) & 1.959(6) \\
    & 0.0212897 & 1.792(7) & 1.795(7) & 1.779(7) & 1.777(6)\\ \\[-10pt]
\hline\\[-10pt]
6.3 & 0.0246230 & 2.042(9) & 2.050(9) & 2.049(9)  & 2.047(9)\\
    & 0.0222430 & 1.953(8) & 1.962(8) & 1.961(8)  & 1.958(7)\\
    & 0.0198558 & 1.830(11)& 1.839(10)& 1.839(10) & 1.833(9)\\
    & 0.0174640 & 1.711(10)& 1.722(9) & 1.720(9)  & 1.720(8) \\ \\[-10pt] 
\Hline\\[-10pt]
\end{tabular}
\end{center}
\caption{The pseudoscalar meson effective masses in the $\pi/2$ case.
The physical kaon in these units is $r_0M^{\rm phys}_K =1.2544$.}
\label{tab:meffpi/2}
\end{table}

\begin{table}[!htp]
\begin{center}
%\scriptsize
\begin{tabular}{cccccc}
\Hline \\[-10pt]
$\beta$ & $a \mu_{0,{\rm tw}}$ & $r_0 F_{sc}$ & $r_0 F_{sd}$ & $r_0 F_{ud}$ &
$r_0 \tilde F_{ud}$ \\ \\[-10pt]
\Hline \\[-10pt]
6.0 & 0.03816 & 0.4279(48) & 0.4543(52) & 0.4833(52) & 0.4818(57) \\
    & 0.03152 & 0.4203(40) & 0.4478(49) & 0.4753(51) & 0.4732(57) \\
    & 0.02708 & 0.4123(39) & 0.4370(43) & 0.4629(43) & 0.4617(49) \\
    & 0.02261 & 0.4006(33) & 0.4269(41) & 0.4495(42) & 0.4491(47) \\ \\[-10pt]
\hline\\[-10pt]
    &   *      & 0.3851(58) & 0.4097(70) & 0.4317(67) & 0.4325(73) \\
\hline \\[-10pt]    
6.1 & 0.0317110 & 0.4653(49) & 0.4795(56) & 0.4939(49) & 0.4936(62) \\
    & 0.0271230 & 0.4417(48) & 0.4559(50) & 0.4731(53) & 0.4716(62) \\
    & 0.0225230 & 0.4306(50) & 0.4437(59) & 0.4549(63) & 0.4536(67) \\ \\[-10pt]
\hline\\[-10pt]
    &    *     & 0.3919(112) & 0.4009(136) & 0.4121(134) & 0.4093(151)
\\  
\hline \\[-10pt]    
6.2 & 0.0283240 & 0.4706(54) & 0.4817(58) & 0.4898(59) & 0.4890(66) \\
    & 0.0259850 & 0.4663(49) & 0.4746(48) & 0.4812(49) & 0.4812(58) \\
    & 0.0212897 & 0.4491(49) & 0.4553(53) & 0.4611(53) & 0.4600(63) \\ \\[-10pt]
\hline\\[-10pt]
    &     *    & 0.4170(144) & 0.4153(158) & 0.4188(156) & 0.4175(182)
\\
\hline \\[-10pt]    
6.3 & 0.0246230 & 0.4693(82) & 0.4781(83) & 0.4856(82) & 0.4847(95) \\
    & 0.0222430 & 0.4632(58) & 0.4695(64) & 0.4764(68) & 0.4758(80) \\
    & 0.0198558 & 0.4581(56) & 0.4637(63) & 0.4688(63) & 0.4691(75) \\
    & 0.0174640 & 0.4380(56) & 0.4450(57) & 0.4517(58) & 0.4501(69) \\ \\[-10pt] 
\hline\\[-10pt]
    &      *   & 0.4080(135) & 0.4115(141) & 0.4166(143) & 0.4151(169) \\ 
\hline\\[-7pt]
    & & {\bf cont.} & {\bf limit } & {\bf extrap.}  & \\ \\[-7pt]
\hline \\[-10pt]
\mbox{all beta}        && 0.430(18) & 0.412(19) & 0.401(19) &
0.396(22) \\[3pt]
\mbox{w/o} $\beta=6.0$  && 0.437(32) & 0.429(35) & 0.424(35) &
0.425(40) \\ \\[-10pt]
\Hline
\end{tabular}
\end{center}
\caption{The pseudoscalar meson decay constants in the $\pi/2$
  case.  Each decay constant value at the physical kaon mass 
  $r_0M_K^{\rm phys} =1.2544$ has been obtained by linear extrapolation in the corresponding
  pseudoscalar mass squared (in $r_0$ units) and is reported in the lines indicated by an asterisk. 
  The continuum limit results are obtained from linear extrapolations in $(a/r_0)^2$.} 
\label{tab:decpi2}
\end{table}

\begin{table}[!htp]
\begin{center}
%\scriptsize
\begin{tabular}{cccccc}
\Hline \\[-10pt]
$\beta$ && $r_0M^{\rm eff}_{sd}$ & $r_0F_{sd}$ & 
$r_0 \tilde M^{\rm eff}_{sd}$ & $r_0 \tilde F_{sd}$  
\\ \\[-10pt]
\Hline \\[-10pt]
6.0 && 1.326(4) &0.3918(33)&1.319(4) & 0.3979(37) \\
    && 1.253(4) &0.3861(33)&1.244(4) & 0.3907(37) \\
    && 1.207(4) &0.3826(33)&1.198(4) & 0.3864(38) \\ \\[-10pt]
\hline\\[-10pt]
    && 1.2544   &0.3864(33)& 1.2544  & 0.3917(37) \\
\hline\\[-10pt]
6.1 && 1.235(6) &0.3938(44)&1.231(5) & 0.4050(54) \\
    && 1.170(6) &0.3895(43)&1.166(5) & 0.4006(54) \\
    && 1.088(6) &0.3844(42)&1.084(6) & 0.3961(54) \\ \\[-10pt]
\hline\\[-10pt]
    && 1.2544   &0.3951(44)& 1.2544   & 0.4064(53) \\ 
\hline
6.2 && 1.299(6) &0.4110(63)&1.295(6) & 0.4032(68) \\
    && 1.182(6) &0.4044(63)&1.176(6) & 0.3943(66) \\ \\[-10pt]
\hline\\[-10pt]
    && 1.2544   &0.4084(62)&1.2544 & 0.4001(65) \\          
\hline\\[-10pt]
6.3 && 1.338(9) &0.4113(66)&1.337(9) & 0.4160(81) \\
    && 1.259(9) &0.4051(63)&1.259(9) & 0.4092(78) \\
    && 1.175(10)&0.3987(59)&1.176(9) & 0.4029(76) \\ \\[-10pt]
\hline\\[-10pt]
    && 1.2544  &0.4048(61)& 1.2544   & 0.4090(75) \\          
\hline\\[-10pt]
6.45 &&2.054(10)&0.4776(66)&2.052(9)& 0.4871(81) \\
     &&1.848(11)&0.4569(65)&1.847(9)& 0.4673(81) \\
     &&1.702(11)&0.4433(63)&1.701(10)&0.4544(82) \\ \\[-10pt]
\hline\\[-10pt]
    && 1.2544  &0.4089(69)& 1.2544  & 0.4216(91) \\          
\hline\\[-7pt]
  & & & {\bf cont.} & {\bf limit } & {\bf extrap.} \\ \\[-7pt]
\hline \\[-10pt]
\mbox{all beta}            &&& 0.420(6)  && 0.424(8) \\    
\mbox{w/o} $\beta=6.0$     &&& 0.419(9) && 0.420(12) \\ \\[-10pt]
\Hline
\end{tabular}
\end{center}
\caption{The pseudoscalar meson effective mass and decay constant in the $\pi/4$
  case. The decay constant values at the physical kaon mass $r_0M_K^{\rm phys} =
  1.2544$ have been obtained by linear interpolation 
  (extrapolation at $\beta = 6.45$) in $r_0M_{sd}^2$. The continuum limit results are
  obtained from linear extrapolations in $(a/r_0)^2$.} 
\label{tab:decpi4}
\end{table}

\begin{table}[!htp]
\begin{center}\scriptsize
\begin{tabular}{cccccc}
\Hline \\[-6pt] 
$\beta$ & $\left[\dfrac{x_0^{\rm min}}{2r_0}, \dfrac{x_0^{\rm max}}{2r_0}\right]$ &
$r_0 M_{sd}$ & $\cR_+^{\rm ALPHA}$ & $\cR_+^{\rm LANL}$  & $\cR_+^{{\rm
  ALPHA;~w/o}~\icA}$ \\ \\[-6pt]
\Hline \\[-6pt]
6.1 & $[1.34,3.08]$ & 2.039(6) & 1.433(11)(10)(15) & 1.403(11)(14)(18)
& 1.339(10)(8)(13) \\ \\[-6pt]
    &               & 1.874(7) & 1.380(13)(10)(16) & 1.351(13)(14)(19) &
1.290(12)(7)(14) \\ \\[-6pt]
    &               & 1.715(6) & 1.332(12)(9)(16) & 1.305(12)(13)(18) &
1.245(11)(7)(13) \\ \\[-6pt]
\hline \\[-6pt]
    & & 1.2544 & 1.219(28)(37) & 1.194(28)(42) & 1.140(25)(31) \\ \\[-6pt]
\hline \hline \\[-6pt]
6.1 & $[1.18,3.55]$ & 1.233(6) & 1.239(14)(9)(16) & 1.204(13)(12)(18) &
1.126(12)(6)(14)
 \\ \\[-6pt]
    & $[1.18,3.55]$ & 1.168(6) & 1.218(15)(9)(17) & 1.183(14)(12)(19) &
1.106(13)(6)(15)    
     \\ \\[-6pt]
    & $[1.26,3.47]$ & 1.086(6) & 1.190(17)(8)(19) & 1.156(16)(12)(20) &
1.080(15)(6)(16)    
     \\ \\[-6pt]
\hline \\[-6pt]
    & & 1.2544 & 1.247(12)(15) & 1.211(12)(16) & 1.134(11)(14)    
     \\ \\[-6pt]
\Hline \\[-6pt]
\end{tabular}
\end{center}
\caption{Results for the pseudoscalar mass and the various ratios 
  ${\cal R}_+$ at $\beta = 6.1$. The upper (lower) part of the table
  corresponds to the $\pi/2$ ($\pi/4$) case. The error in $r_0 M_{sd}$ is
  statistical. The three errors of each ${\cal R}_+$ ratio are,
  in order of appearance:  (i) due to the statistical
  fluctuations of the correlations; (ii) due to the errors of 
  $\ZA$ and $\ZV$; (iii) the total error from the two previous ones. The results of 
  the extrapolations to the physical kaon mass value
  are shown at the bottom of
  each $\beta$-dataset: the first error of the extrapolated ${\cal R}_+$ 
  is that arising from type-(i) errors of the fitted values, while the
  second is from type-(iii) errors.}
\vskip 0.2cm
\label{tab:resR+}
\end{table}

\begin{table}[!htp]
\begin{center}
\begin{tabular}{ccc}
\Hline \\[-10pt] 
$\beta$ & $\hat B_K(\alpha=\pi/2)$ & $\hat B_K(\alpha=\pi/4)$  \\ \\[-10pt]
\Hline \\[-10pt]
6.0        & 0.911(22)       & 0.926(13) \\
{\bf 6.1}  & {\bf 0.824(26)} & {\bf 0.843(13)} \\ 
6.1        & {0.812(24)}     & {0.779(13)} \\
6.2        & 0.828(30)       & 0.798(14) \\
6.3        & 0.789(34)       & 0.775(15) \\ 
6.45       & -- \ \          & 0.789(19) \\ \\[-10pt]
\Hline \\[-10pt]
\end{tabular}
\end{center}
\caption{The RGI bag parameter $\hat B_K$ for all lattice spacings.
The second (third) column displays data from  $\alpha=\pi/2$ ($\alpha=\pi/4$) tmQCD.
All results are from ref.~\cite{Dimopoulos:2006dm}, except those at
$\beta = 6.1$ in boldface, which have been obtained in the
present work. The errors, which incorporate both
statistical and systematic effects, have been estimated as 
described in
ref.~\cite{Dimopoulos:2006dm}.}
\vskip 0.2cm
\label{tab:BK}
\end{table}

\begin{table}[!htp]
\begin{center}\scriptsize
\begin{tabular}{cccccc}
\Hline \\[-6pt] 
$\beta$ & $\left[ \dfrac{x_0^{\rm min}}{2r_0}, \dfrac{x_0^{\rm max}}{2r_0} \right]$ &
$r_0 M_{sd}$ & $\cR_-^{\rm ALPHA}$ & $\cR_-^{\rm LANL}$  & $\cR_-^{{\rm
  ALPHA;~w/o}~\icA}$ \\
&&&&& \\[-6pt]
\Hline \\[-6pt]
6.0 & $[1.30,3.17]$ & 2.092(6) & 2.632(32)(19)(37) & 2.435(30)(25)(39) &
2.293(28)(13)(31)
 \\ \\[-6pt]
    &               & 1.907(7) & 2.732(43)(19)(47) & 2.528(40)(26)(48) &
2.381(38)(14)(40)
 \\ \\[-6pt]
    &               & 1.780(6) & 2.809(45)(20)(49) & 2.596(41)(27)(50) &
2.443(40)(14)(42)
 \\ \\[-6pt]
    &               & 1.635(6) & 2.852(46)(20)(50) & 2.635(42)(28)(51) &
2.480(40)(14)(43)
 \\ \\[-6pt]
\hline \\[-6pt]
    & & 1.2544 & 3.009(66)(73) & 2.778(61)(75) & 2.614(58)(62) \\ \\[-6pt]
\hline \\[-6pt]
6.1 & $[1.34,3.08]$ & 2.039(6) & 2.450(27)(17)(32) & 2.400(27)(25)(36) &
2.289(26)(13)(29)
 \\ \\[-6pt]
    &               & 1.874(7) & 2.627(32)(19)(37) & 2.574(32)(26)(41) &
2.457(30)(14)(33)
 \\ \\[-6pt]
    &               & 1.715(6) & 2.697(40)(19)(44) & 2.641(39)(27)(48) &
2.521(38)(14)(40)
 \\ \\[-6pt]
\hline \\[-6pt]
    & & 1.2544 & 3.010(83)(94) & 2.95(8)(10) & 2.817(79)(85) \\ \\[-6pt]
\hline \\[-6pt]
6.2 & $[1.29,3.05]$ & 2.079(6) & 2.414(28)(17)(32) & 2.400(27)(25)(37) &
2.312(27)(13)(30)
 \\ \\[-6pt]
    &               & 1.980(7) & 2.449(30)(17)(34) & 2.435(30)(25)(39) &
2.347(29)(13)(32)
 \\ \\[-6pt]
    &               & 1.795(7) & 2.623(39)(19)(43) & 2.608(39)(27)(47) &
2.513(38)(14)(40)
 \\ \\[-6pt]
\hline \\[-6pt]
    & & 1.2544 & 2.92(10)(12) & 2.91(10)(13) & 2.80(10)(11) \\ \\[-6pt]
\hline \\[-6pt]
6.3 & $[1.23,3.00]$ & 2.050(9) & 2.352(39)(17)(43) & 2.366(39)(24)(46) &
2.284(38)(13)(40)
 \\ \\[-6pt]
    &               & 1.962(8) & 2.506(45)(18)(49) & 2.522(45)(26)(52) &
2.433(44)(14)(46)
 \\ \\[-6pt]
    &               & 1.839(10) & 2.489(56)(18)(59) & 2.504(56)(26)(62) &
2.417(55)(14)(56)
 \\ \\[-6pt]
    &               & 1.722(9) & 2.661(58)(19)(61) & 2.677(58)(27)(64) &
2.584(56)(15)(58)
 \\ \\[-6pt]
\hline \\[-6pt]
    & & 1.2544 & 2.95(12)(12) & 2.96(12)(13) &
2.85(11)(12) \\ \\[-6pt]
\Hline \\[-6pt]
\end{tabular}
\end{center}
\caption{Results for the pseudoscalar mass and the various ratios 
  ${\cal R}_-$ for twist angle $\pi/2$. For an explanation of the errors,
  see the caption of Table~\ref{tab:resR+}.
}
\vskip 0.2cm
\label{tab:respi2_minus}
\end{table}

\begin{table}[!htp]
\begin{center}\scriptsize
\begin{tabular}{cccccccc}
\Hline \\[-6pt] 
$\beta$ & $\left[ \dfrac{x_0^{\rm min}}{2r_0}, \dfrac{x_0^{\rm max}}{2r_0} 
\right]$ &
$r_0 M_{sd}$ & $\cR_-^{\rm ALPHA}$ & $\cR_-^{\rm LANL}$  & $\cR_-^{{\rm
  ALPHA;~w/o}~\icA}$ \\ \\[-6pt]
\Hline \\[-6pt]
6.0 & $[1.30,3.17]$ & 1.326(4) &
3.202(40)(23)(46) & 2.880(37)(31)(48) & 2.657(34)(16)(38)
\\ \\[-6pt]
    & $[1.30,3.17]$ & 1.253(4) & 3.306(47)(23)(53) & 2.971(42)(32)(53) &
2.739(40)(16)(43)
\\ \\[-6pt]
    & $[1.30,3.17]$ & 1.207(4) & 3.379(54)(24)(59) & 3.035(48)(33)(58) &
2.798(45)(17)(48)
 \\ \\[-6pt]
\hline \\[-6pt]
    & & 1.2544 & 3.304(46)(53) & 2.968(41)(53) & 2.738(38)(41)
     \\ \\[-6pt]
\hline \\[-6pt]
6.1 & $[1.18,3.55]$ & 1.233(6) & 3.199(67)(23)(71) & 3.112(65)(32)(73) &
2.909(62)(17)(64)
 \\ \\[-6pt]
    & $[1.18,3.55]$ & 1.168(6) &
3.296(77)(23)(80) & 3.204(74)(33)(81) & 2.995(70)(17)(72)
     \\ \\[-6pt]
    & $[1.26,3.47]$ & 1.086(6) &
3.429(91)(24)(94) & 3.330(88)(35)(95) & 3.112(83)(18)(85)
     \\ \\[-6pt]
\hline \\[-6pt]
    & & 1.2544 & 3.160(62)(65) & 3.077(60)(68) & 2.875(57)(59)
     \\ \\[-6pt]
\hline \\[-6pt]
6.2 & $[1.35,3.52]$ & 1.299(6) & 2.999(58)(21)(61) & 2.976(57)(32)(65) &
2.828(55)(16)(57)
 \\ \\[-6pt]
    & $[1.35,3.52]$ & 1.182(6) &
3.127(72)(22)(75) & 3.102(72)(33)(79) & 2.945(68)(17)(70)
     \\ \\[-6pt]
\hline \\[-6pt]
    & & 1.2544 &
3.050(62)(65) & 3.026(61)(68) & 2.875(58)(60)
     \\ \\[-6pt]
\hline \\[-6pt]
6.3 & $[1.29,2.94]$ & 1.338(9) & 2.964(64)(21)(68) & 2.983(64)(31)(71) &
2.848(62)(16)(64)
 \\ \\[-6pt]
    & $[1.35,2.88]$ & 1.259(9) & 3.056(73)(22)(76) & 3.081(73)(32)(80) &
2.936(70)(17)(72)
     \\ \\[-6pt]
    & $[1.47,2.76]$ & 1.175(10) & 3.164(85)(22)(88) & 3.204(85)(33)(91) &
3.040(82)(17)(84)
    \\ \\[-6pt]
\hline \\[-6pt]
    & & 1.2544 & 3.065(72)(76) & 3.093(72)(79) &
2.945(69)(71)
     \\ \\[-6pt]
\hline \\[-6pt]
6.45& $[1.30,2.84]$ & 2.054(10) & 2.262(41)(16)(44) & 2.285(41)(23)(47) &
2.203(40(12)(42)
\\ \\[-6pt]
    & $[1.25,2.88]$ & 1.848(11) & 2.384(53)(17)(56) & 2.408(53)(24)(58) &
2.322(52)(13)(54)
     \\ \\[-6pt]
    & $[1.30,2.84]$ & 1.702(11) & 2.484(66)(18)(68) & 2.508(66)(25)(70) &
2.419(64)(13)(66)
     \\ \\[-6pt]
\hline \\[-6pt]
    & & 1.2544 & 
2.700(93)(98) & 2.725(93)(99) & 2.630(90)(93)
     \\ \\[-6pt]
\Hline
\end{tabular}
\end{center}
\caption{Results for the pseudoscalar mass and the various ratios 
  ${\cal R}_-$ for twist angle $\pi/4$.
  For an explanation of the errors, see the caption of Table~\ref{tab:resR+}.
}
\label{tab:respi4_minus}
\end{table}
 
\begin{table}[!htp]
\begin{center}
\begin{tabular}{ccc}
\Hline \\[-10pt] 
$a\mu_{0,{\rm tw}}$ & $\cot(\taa)$ & $\cot(\taa)$ \\ \\[-10pt]
& \req{eq:tanapcac} & \req{eq:tanapcvc} \\ \\[-10pt]
\Hline \\[-10pt]
0.0317110 & 0.102(2) & 0.103(2) \\
0.0271230 & 0.141(3) & 0.143(3) \\
0.0225230 & 0.188(3) & 0.191(3) \\ \\[-10pt]
\hline\hline \\[-10pt]
0.00810 & 1.435(6) & 1.413(6) \\
0.00720 & 1.495(7) & 1.473(6) \\
0.00615 & 1.583(8) & 1.561(8) \\ \\[-10pt]
\Hline \\[-10pt]
\end{tabular}
\end{center}
\caption{Twist angles, computed with an offset critical hopping parameter
  $\kappa_{\rm cr}^\prime(\beta = 6.1)$ and from two different procedures, 
  indicated by the corresponding equations. The upper (lower) part of the 
  table corresponds to the $\pi/2$ ($\pi/4$) case. }
\label{tab:cota61}
\end{table}

\begin{table}[!htp]
\begin{center}
\begin{tabular}{cccccccc}
\Hline \\[-10pt] 
$a\mu_{0,{\rm tw}}$ & $r_0 M^{\rm eff}_{sd}$ & $r_0 M^{\rm eff}_{ud}$ &
$r_0 \tilde M^{\rm eff}_{ud}$ & $r_0 F_{sd}$ &  $r_0 F_{ud}$ &
$r_0 \tilde F_{ud}$\\ \\ [-10pt]
\Hline \\[-10pt]
0.0317110 & 1.979(6) & 1.910(6) & 1.904(6) & 0.4697(55) & 
0.4745(55) & 0.4746(60) \\
0.0271230 & 1.812(7) & 1.739(7) & 1.735(6) & 0.4462(48) &
0.4509(49) & 0.4502(57) \\
0.0225230 & 1.647(7) & 1.575(6) & 1.568(6) & 0.4311(55) &
0.4284(56) & 0.4284(60) \\ \\[-10pt]
\Hline
\end{tabular}
\end{center}
\caption{Pseudoscalar effective masses and decay constants, for the
  $\pi/2$ case, computed with the offset critical hopping parameter
  $\kappa_{\rm cr}^\prime(\beta = 6.1)$.
}
\label{tab:mf61pi2}
\end{table}

\begin{table}[!htp]
\begin{center}
\begin{tabular}{ccccc}
\Hline \\[-10pt] 
$\beta$ & $a\mu_{0,{\rm tw}}$ & 
$\left(M^{\rm eff}_{sd}/M^{\rm eff}_{sc}\right)^2  -1$ &
$\left(M^{\rm eff}_{ud}/M^{\rm eff}_{sc}\right)^2  -1$ &
$\Zm (\delta a m_{\rm cr}) / a \mren{,{\rm w}}$ \\ \\[-10pt]
\Hline \\[-10pt]
6.1 & 0.0317110 & -0.044 (1) & -0.109 (2) & -0.06 \\
    & 0.0271230 & -0.050 (1) & -0.124 (3) & -0.07 \\
    & 0.0225230 & -0.053 (2) & -0.134 (4) & -0.09 \\ \\[-10pt]
\Hline \\[10pt]
\end{tabular}
\end{center}
\vskip -1.0cm
\caption{
Comparison of the flavour breaking effects in the pseudoscalar masses,
computed with the offset $\kappa_{\rm cr}^\prime(\beta = 6.1)$. 
In the last column, we also show the results of the ``prediction'' based on
\req{eq:massratFB}.
}
\vskip 0.2cm
\label{tab:respi2ratios}
\end{table}

\begin{table}[!htp]
\begin{center}
\begin{tabular}{ccccc}
\Hline \\[-10pt] 
$a\mu_{0,{\rm tw}}$ & $r_0 M^{\rm eff}_{sd}$ & $r_0 \tilde M^{\rm eff}_{sd}$ &
$r_0 F_{sd}$ & $r_0 \tilde F_{sd}$\\ \\ [-10pt]
\Hline \\[-10pt]
0.00810 & 1.382(5) & 1.372(5) & 0.4001(36) & 0.3325(34) \\
0.00720 & 1.326(5) & 1.316(5) & 0.3944(36) & 0.3196(33) \\
0.00615 & 1.257(5) & 1.247(5) & 0.3862(36) & 0.3025(33) \\ \\[-10pt]
\Hline
\end{tabular}
\end{center}
\caption{Pseudoscalar effective masses and decay constants, for the
  $\pi/4$ case, computed with the offset $\kappa_{\rm cr}^\prime(\beta = 6.1)$.
}
\label{tab:mf61pi4}
\end{table}
\vskip 0.3cm
\begin{table}[!htp]
\begin{center}
\begin{tabular}{cccc}
\Hline \\[-10pt] 
$\beta$ & $a\mu_{0,{\rm tw}}$ & 
$\left(\bar M^{\rm eff}_{sd}/M^{\rm eff}_{sd}\right)^2  -1$ &
$\Zm (\delta a m_{\rm cr}) / 2 a \muren{,{\rm w}}$ \\ \\[-10pt]
\Hline  \\[-10pt]
6.1 & 0.00810 & 0.25 (2) & 0.25 \\
    & 0.00720 & 0.28 (2) & 0.28 \\
    & 0.00615 & 0.34 (2) & 0.33 \\ \\[-10pt]
\Hline \\[-10pt]
\end{tabular}
\end{center}
\caption{Comparison of the effect of $\kappa_{\rm cr}'(\beta=6.1)$ on the  
  the pseudoscalar mass, measured in the simulation, with the ``prediction'' based on
  \req{eq:massratFB4}.}
\label{tab:respi4ratios}
\end{table}

\newpage
\end{appendix}

\bibliography{lattice}        %or whatever

\end{document}